\documentclass[10pt,aps,pra,twocolumn]{revtex4-1}

\usepackage[dvips]{graphicx}
\usepackage{epsfig}

\usepackage[margin=2cm,head=0.5cm]{geometry}
\usepackage{bm}
\usepackage{amsmath}
\usepackage{amssymb}
\usepackage{latexsym}
\usepackage{amsfonts}
\usepackage{epsfig}
\usepackage{color}
\usepackage[linktocpage, colorlinks=true ,linkcolor=blue, citecolor=blue]{hyperref}
\usepackage[all]{hypcap}
\usepackage[format=plain,
      justification=RaggedRight,
      singlelinecheck=false]
     {caption}

\newcommand{\nn}{\nonumber\\}

\newcommand{\f}[1]{\mbox{\boldmath$#1$}}

\newcommand{\bea}{\begin{eqnarray}}
\newcommand{\eea}{\end{eqnarray}}
\newcommand{\beann}{\begin{eqnarray*}}
\newcommand{\eeann}{\end{eqnarray*}}
\newcommand{\ord}{{\cal O}}

\newcommand{\abs}[1]{{\left| #1 \right|}}
\newcommand{\res}[1]{\underset{ #1 }{\rm Res}}

\newcommand{\ii}{\mathrm{i}}  % alternativ: {$\mathbf{i}$}  % imgaginäre Einheit

\begin{document}
  
\title{Transmission from reverse reaction coordinate mappings}

\author{Niklas Martensen}
\author{Gernot Schaller}
\email{gernot.schaller@tu-berlin.de}
\affiliation{Institut f\"ur Theoretische Physik, Hardenbergstr. 36, Technische Universit\"at Berlin, D-10623 Berlin, Germany}
\date{\today}

\begin{abstract}
We point out that the transport properties of non-interacting fermionic chains tunnel-coupled to two reservoirs at their ends can be mapped to
those of a single quantum dot that is tunnel-coupled to two transformed reservoirs.
The parameters of the chain are mapped to additional structure in the spectral densities of the transformed reservoirs.
For example, this enables the calculation of the transmission of quantum dot chains by evaluating the known transmission
of a single quantum dot together with structured spectral densities.
We exemplify this analytically for short chains,
which allows to optimize the transmission.
In addition, we also demonstrate that the mapping can be performed numerically by computing
the transmission of a Su-Schrieffer-Heeger chain.
\end{abstract}

\maketitle

%%%%%%%%%%%%%%%%%%%%%%%%%%%%%%%%%%%%%%%%%%%%%%%%%%%%%%%%%%%%%%%%%%%%%%%%%%%%%%%%%%%%%%%%%%%%%%%%%%%
%%%%%%%%%%%%%%%%%%%%%%%%%%%%%%%%%%%%%%%%%%%%%%%%%%%%%%%%%%%%%%%%%%%%%%%%%%%%%%%%%%%%%%%%%%%%%%%%%%%
%%%%%%%%%%%%%%%%%%%%%%%%%%%%%%%%%%%%%%%%%%%%%%%%%%%%%%%%%%%%%%%%%%%%%%%%%%%%%%%%%%%%%%%%%%%%%%%%%%%
%%%%%%%%%%%%%%%%%%%%%%%%%%%%%%%%%%%%%%%%%%%%%%%%%%%%%%%%%%%%%%%%%%%%%%%%%%%%%%%%%%%%%%%%%%%%%%%%%%%
\section{Introduction}

The significant advances in time-resolved charge detectors have allowed to investigate the 
Full Counting Statistics of electrons in great detail~\cite{flindt2009a}.
From the theoretical perspective, perturbative approaches such as quantum master equations~\cite{schaller2009b} suffer from quite rigid 
constraints on the allowed parameter regimes within which they are valid.
To the contrary, non-interacting electronic systems can be solved exactly.
In electronic counting statistics, the famous Levitov-Lesovik 
formula \cite{levitov1993a,klich2003a,schoenhammer2007a} connects
the cumulant-generating function of the complete electronic counting statistics
of general non-interacting two-terminal systems with an electronic transmission probability.
Although having a very intuitive interpretation, the transmission probability in general needs to be calculated from non-equilibrium Greens functions~\cite{haug2008,economou2006}.
These in turn have to be computed via matrix inversion methods from the free Greens function,
which is analytically known for particularly shaped reservoirs with particular spectral densities only.
On the other hand, for a single quantum dot coupled to two fermionic leads, the transmission function can be calculated exactly and does not rely on any assumption on the spectral density~\cite{topp2015a}.
Therefore, it may appear intuitively possible to map a chain-type system with structureless fermionic reservoirs to
a shorter chain-type system coupled to structured reservoirs at its end.
In the literature, such mapping relations are well-known for particular examples~\cite{zedler2009a}.

The reaction-coordinate mapping is a well-known technique for bosonic systems~\cite{martinazzo2011a,woods2014a,strasberg2016a}, which shifts the boundary between system and
its reservoirs by transferring suitable reservoir degrees of freedom into the system. 
The advantage of this procedure is that the transformed model may capture strong-coupling and non-Markovian effects of the original system by a perturbative treatment of the re-defined system-reservoir coupling, at the expense of increasing the dimension of the problem.
The method can also be extended to fermionic models~\cite{schaller2018a,strasberg2018a,nazir2018a} and turned out particularly useful for identifying thermodynamic quantities.
However, it is also obvious that the approach must fail when the redefined system-reservoir coupling does not allow for a perturbative treatment.

Fortunately, the mapping techniques are independent of the treatment applied afterwards.
They can be equally well used in combination with formally exact 
nonequilibrium Greens function techniques, which allow to compute global observables of system and reservoir.
While in this context, a redefinition of the system-reservoir boundary does not extend the regime of
validity, it may nevertheless be helpful by e.g. decreasing the dimension of the central system, reducing the numerical effort in obtaining the Greens function.
In addition, we find that such mapping relations may help in developing some intuition to optimize microscopic system parameters for a given objective.
Specifically, we will point out that for fermionic chain models, 
the sought-after mapping relations can be easily found by
inverting the fermionic reaction-coordinate mapping.

This paper is organized as follows:
In Sec.~\ref{SEC:transmission} we motivate the problem by introducing basic concepts of electronic transport, discussing the Levitov-Lesovik formula
and the simple example of the single-electron transistor.
Afterwards, we introduce the reverse reaction-coordinate mapping in Sec.~\ref{SEC:reverse_mapping}.
We exemplify this for particular chains both analytically and numerically in Sec.~\ref{SEC:examples}
before concluding.
Appendices provide details on the pole structure, the non-equilibrium Greens function technique, and  on the used numerical methods.

%%%%%%%%%%%%%%%%%%%%%%%%%%%%%%%%%%%%%%%%%%%%%%%%%%%%%%%%%%%%%%%%%%%%%%%%%%%%%%%%%%%%%%%%%%%%%%%%%%%
%%%%%%%%%%%%%%%%%%%%%%%%%%%%%%%%%%%%%%%%%%%%%%%%%%%%%%%%%%%%%%%%%%%%%%%%%%%%%%%%%%%%%%%%%%%%%%%%%%%
%%%%%%%%%%%%%%%%%%%%%%%%%%%%%%%%%%%%%%%%%%%%%%%%%%%%%%%%%%%%%%%%%%%%%%%%%%%%%%%%%%%%%%%%%%%%%%%%%%%
%%%%%%%%%%%%%%%%%%%%%%%%%%%%%%%%%%%%%%%%%%%%%%%%%%%%%%%%%%%%%%%%%%%%%%%%%%%%%%%%%%%%%%%%%%%%%%%%%%%
\section{Transmission}\label{SEC:transmission}

In this section, we first discuss the relevance of the transmission in the Levitov-Lesovik formula 
and then make it explicit for a single quantum dot coupled to two reservoirs.

%%%%%%%%%%%%%%%%%%%%%%%%%%%%%%%%%%%%%%%%%%%%%%%%%%%%%%%%%%%%%%%%%%%%%%%%%%%%%%%%%%%%%%%%%%%%%%%%%%%

\subsection{Levitov-Lesovik formula}
For non-interacting electronic transport between two terminals connected via a central system, 
the long-term cumulant-generating function for the distribution of electrons transferred from left to right can be compactly 
expressed via the Levitov-Lesovik formula~\cite{levitov1993a,levitov1996a,klich2003a}
\begin{align}\label{EQ:cgf}
C(\chi) = \int \frac{d\omega}{2\pi} \ln \Big[1+ T(\omega) \big(f_{LR}(\omega)&(e^{+\ii\chi}-1) \nn
+ f_{RL}(\omega)&(e^{-\ii\chi}-1)\big)\Big]\, .
\end{align}
Here, $0\le T(\omega) \le 1$ denotes the transmission probability for an electron to pass the system at energy $\omega$,
$f_\alpha(\omega)=[e^{\beta_\alpha (\omega-\mu_\alpha)}+1]^{-1}$ denotes the Fermi function of lead $\alpha\in\{L,R\}$, characterized
by inverse temperature $\beta_\alpha$ and chemical potential $\mu_\alpha$ and $f_{\alpha\bar\alpha}(\omega)=f_\alpha(\omega)\left[1-f_{\bar\alpha}(\omega)\right]$.
In particular, we obtain the current by computing the first derivative with respect to the counting field
$I=-\ii \partial_\chi C(\chi)|_{\chi=0}$, which yields the Landauer formula~\cite{landauer1957a}.
Similarly, the noise is computed by the second derivative\\
\mbox{$S = - \partial_\chi^2 C(\chi)|_{\chi=0}$}.
Explicitly, these read
\begin{align}\label{EQ:current_noise}
I = \int & \frac{d\omega}{2\pi} T(\omega) \left[f_L(\omega) - f_R(\omega)\right]\,,\nn
S = \int & \frac{d\omega}{2\pi} \Big\{T(\omega) \left[f_{LL}(\omega)+f_{RR}(\omega)\right]\nn
&+ T(\omega)[1-T(\omega)]\left[f_L(\omega) - f_R(\omega)\right]^2\Big\}\,.
\end{align}
This underlines the importance of the transmission $T(\omega)$ for transport quantities.
It can be obtained with the help of the retarded central Greens function of the central system~\cite{economou2006,haug2008} 
\begin{align}
T(\omega) = \abs{G_{1,N}(\omega)}^2 \Gamma_L(\omega) \Gamma_R(\omega)\,,
\end{align}
where $\Gamma_\alpha(\omega) = 2\pi \sum_k \abs{t_{k\alpha}}^2 \delta(\omega-\epsilon_{k\alpha})$ 
denote the electronic tunneling rates.
The central Greens function $G(\omega)$ in turn has to be determined numerically by inversion of a matrix, such that the actual computation of the
transmission can be demanding, in particular for transport through long chains~\cite{boehling2018a}.
In addition, to control or engineer the transmission $T(\omega)$ with respect to some given objective, an analytic form of it would be desirable. 

To illustrate the potential use of an optimized transmission, consider the case where formally the transmission function has a sharp transmission window, i.e., is a box-shaped function with $T(\omega)=1$ in an 
interval $[\omega_{\rm min},\omega_{\rm max}]$ and $T(\omega)=0$ elsewhere.
Then, one could apply a large bias voltage
such that $f_L(\omega)=1$ and $f_R(\omega)=0$ for all $\omega\in[\omega_{\rm min},\omega_{\rm max}]$.
This just means that the transport window (defined at low temperatures by the difference of the chemical potentials) 
includes the transmission window.
Then, the integral in~(\ref{EQ:cgf}) collapses 
$C(\chi)\to \ii \frac{\omega_{\rm max}-\omega_{\rm min}}{2\pi} \chi$, which would realize noiseless transport
in a far-from equilibrium situation.
This situation is certainly over-idealized but illustrates that by controlling transmission, interesting applications come within reach.
Even when the transmission is not perfectly box-shaped, the noise can be lowered by its shape.
This allows one to study quantum violations~\cite{agarwalla2018a} of the
thermodynamic uncertainty relation known for Markovian rate equations obeying  detailed balance~\cite{barato2015a,pietzonka2016a,gingrich2016a,horowitz2017a,pietzonka2018a}.
In addition, we mention here that devices with a structured transmission function can be used as energy-resolving charge detectors similar to 
quantum point contacts, providing also information on the energy transfer between the probed system and the charge detector.

%%%%%%%%%%%%%%%%%%%%%%%%%%%%%%%%%%%%%%%%%%%%%%%%%%%%%%%%%%%%%%%%%%%%%%%%%%%%%%%%%%%%%%%%%%%%%%%%%%%

\subsection{Single electron transistor}

The transmission of a single quantum dot with energy $\epsilon$ coupled to two structured leads with energy-dependent spectral 
coupling densities $\Gamma_{L/R}(\omega)$ 
is well known~\cite{haug2008,topp2015a}
\begin{align}\label{EQ:transmission}
T(\omega) = \frac{\Gamma_L(\omega) \Gamma_R(\omega)}{\left[\omega-\epsilon-\Sigma(\omega)\right]^2
+ \left[\frac{\Gamma_L(\omega) + \Gamma_R(\omega)}{2}\right]^2}\,.
\end{align}
Here, the level renormalization induced by the coupling to the leads is given by
\begin{align}\label{EQ:level_renormalization}
\Sigma(\omega) = \frac{1}{2\pi} \sum_\alpha {\cal P} \int \frac{\Gamma_\alpha(\omega')}{\omega-\omega'} d\omega'\,,
\end{align}
where ${\cal P}$ denotes the Cauchy principal value.
Thereby, the transmission $T(\omega)$ is fully defined by the tunneling rates $\Gamma_\alpha(\omega)$ and the dot energy $\epsilon$.
In the wideband limit $\Gamma_\alpha(\omega) \to \Gamma_\alpha$, the level renormalization vanishes and the transmission becomes a simple Lorentzian function, with a width proportional to the coupling strength to the reservoirs, exemplifying the level broadening due to the reservoir-coupling.
In contrast, when $\Gamma_\alpha(\omega)$ has a peak at some frequency $\omega_\alpha$ and is smooth elsewhere, the level renormalization integral will
no longer vanish and will thus effectively modify the behaviour of the transmission.
By choosing $\Gamma_\alpha(\omega) = \Gamma \sqrt{1-(\omega-\epsilon)^2 / \Gamma^2}$ for 
$\epsilon-\Gamma < \omega < \epsilon+\Gamma$, one can show
that in this interval $T(\omega)=1$ and outside this interval $T(\omega)=0$.
Such semicircle spectral densities can be implemented by homogeneous semi-infinite tight-binding chains~\cite{bulnes_cuetara2013a} and thus require carefully fine-tuned reservoirs.

In particular within the context of quantum heat engines, it may be desirable to control the shape of the transmission.
For example, to obtain a tight-coupling regime one may want the transmission to be highly peaked around the central dot energy $\epsilon$ and to be as large as possible $T(\epsilon)=1$.
For the self-energy, one way to fulfill this would be to choose $\Sigma(\epsilon)=0$ and at the same time $\Gamma_L(\epsilon)=\Gamma_R(\epsilon)$.
In addition, the tunneling rates should not generate additional undesired peaks.
An alternative objective could be to generate a transmission that is one around a rather large region centered around $\epsilon$.
This could be achieved by $\Gamma_L(\omega)  \approx \Gamma_R(\omega)$ and $\omega-\epsilon-\Sigma(\omega) \approx 0$ in that region.
In order to deduce the microscopic parameters to generate such situations, some analytic understanding of the tunneling rates entering
the transmission is therefore beneficial.

%%%%%%%%%%%%%%%%%%%%%%%%%%%%%%%%%%%%%%%%%%%%%%%%%%%%%%%%%%%%%%%%%%%%%%%%%%%%%%%%%%%%%%%%%%%%%%%%%%%
%%%%%%%%%%%%%%%%%%%%%%%%%%%%%%%%%%%%%%%%%%%%%%%%%%%%%%%%%%%%%%%%%%%%%%%%%%%%%%%%%%%%%%%%%%%%%%%%%%%
%%%%%%%%%%%%%%%%%%%%%%%%%%%%%%%%%%%%%%%%%%%%%%%%%%%%%%%%%%%%%%%%%%%%%%%%%%%%%%%%%%%%%%%%%%%%%%%%%%%
%%%%%%%%%%%%%%%%%%%%%%%%%%%%%%%%%%%%%%%%%%%%%%%%%%%%%%%%%%%%%%%%%%%%%%%%%%%%%%%%%%%%%%%%%%%%%%%%%%%
\section{Reverse RC mapping and two-terminal application}\label{SEC:reverse_mapping}

In this section, we will suggest that by mapping electronic tight-binding chains with reservoirs to a single quantum dot with modified reservoirs,
one can obtain the transmission of the setup in a simple fashion.
The mapping we will use is conceptionally based on earlier results for bosonic systems~\cite{burkey1984a,garg1985a,martinazzo2011a,woods2014a}, but 
it can be extended to fermionic systems in a straightforward way~\cite{strasberg2018a,schaller2018a,nazir2018a}.
Typically, such mappings are used to transfer degrees of freedom from the reservoir into the system
to allow a perturbative treatment in a re-defined system-reservoir boundary.
What is new in the present work is that we suggest to apply these mappings in a reversed way, effectively transferring degrees of freedom from the system to the reservoir.
Ideally, this would transform a chain connected to two reservoirs at its ends via simple tunneling rates $\Gamma_\alpha(\omega)$ 
to a single quantum dot connected to two reservoirs via more structured tunneling rates $\tilde{\Gamma}_\alpha(\omega)$.
These transformed, more structured tunneling rates can be plugged into the transmission formula~(\ref{EQ:transmission}) for a single quantum dot
and thereby allow to deduce the complete counting statistics of non-trivial chain models.

%%%%%%%%%%%%%%%%%%%%%%%%%%%%%%%%%%%%%%%%%%%%%%%%%%%%%%%%%%%%%%%%%%%%%%%%%%%%%%%%%%%%%%%%%%%%%%%%%%%

\subsection{Single reservoir mapping}

\begin{figure}[ht]
\includegraphics[width=0.45\textwidth,clip=true]{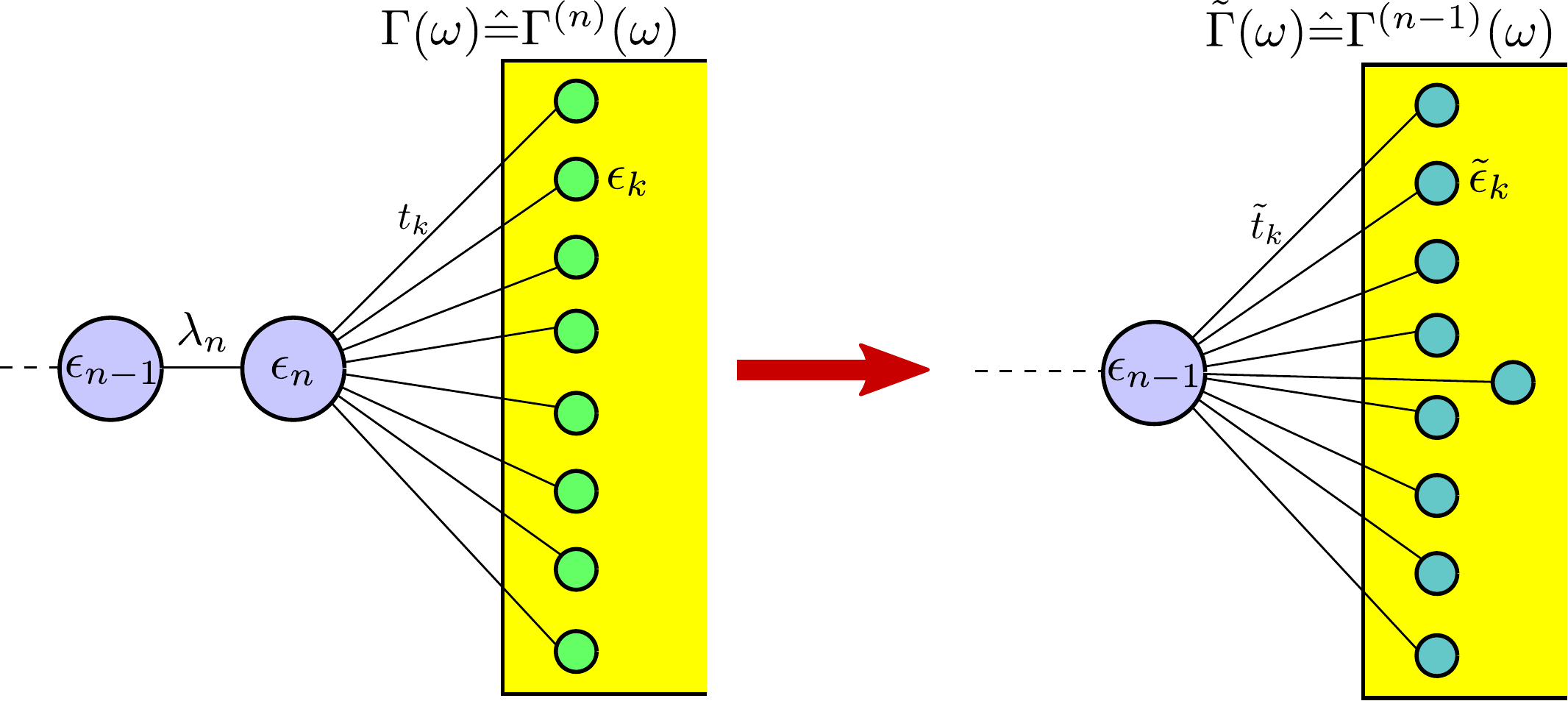}
\caption{\label{FIG:mapping_sketch}
Sketch of the reverse reaction coordinate mapping, translating the boundary mode of a chain (blue) into the reservoir (yellow), effectively reducing the chain length by one.
To obtain the transformed tunneling rate, the corresponding Bogoliubov transform need not be worked out explicitly, but a mapping relation~(\ref{EQ:mapping}) exists that can be evaluated with functional calculus or numerically.
}
\end{figure}

The reverse reaction coordinate mapping can be straightforwardly obtained from results in the literature, such that here we just sketch the essential steps.
We postulate the existence of a reservoir Bogoliubov transform~\cite{woods2014a,nazir2018a} mapping a Hamiltonian of the form
\begin{align}\label{EQ:ham1}
H = & H_{\rm rest} + \left( \lambda d_b^\dagger d + \mathrm{h.c.}\right) + \epsilon d_b^\dagger d_b\nn
& + \left(d_b^\dagger \sum_k t_k c_k + \mathrm{h.c.}\right) + \sum_k \epsilon_k c_k^\dagger c_k\,,
\end{align}
with some (otherwise arbitrary) Hamiltonian $H_{\rm rest}$ that obeys $[H_{\rm rest}, d_b] = 0 = [H_{\rm rest}, c_k]$, system-boundary-mode 
tunneling amplitude $\lambda>0$ (phases can be absorbed in the definition of operators for 1d chains), boundary-mode on-site energy $\epsilon$,  
boundary-mode reservoir tunneling amplitude $t_k$, and reservoir energies $\epsilon_k$,
to another representation, where
\begin{align}\label{EQ:ham2}
H = H_{\rm rest} 
+\left( d^\dagger \tilde{\sum_k} \tilde{t}_k \tilde{c}_k + \mathrm{h.c.}\right)
+ \tilde{\sum_k}  \tilde\epsilon_k \tilde{c}_k^\dagger \tilde{c}_k \, ,
\end{align}
characterized by renormalized tunneling amplitudes $\tilde{t}_k$ and energies $\tilde\epsilon_k$ of a re-defined reservoir.
The Bogoliubov transform involves only the boundary mode $d_b$ and the reservoir modes $c_k$ and therefore, the resulting transformed reservoir 
with modes $\tilde{c}_k$ has one mode more than the original which we denote by the $\,\tilde{\phantom{c}}\,$ symbols over the summation signs, compare also Fig.~\ref{FIG:mapping_sketch}.
We will however be interested in the continuum limit of infinitely many and dense reservoir modes, where
we can define spectral densities (tunnelling rates) in the standard way
\begin{align}
\Gamma(\omega) &= 2\pi \sum_k \abs{t_k}^2 \delta(\omega-\epsilon_k)\,,\nn
\tilde{\Gamma}(\omega) &= 2\pi \sum_k \abs{\tilde{t}_k}^2 \delta(\omega-\tilde{\epsilon}_k)\,.
\end{align} 
In the continuum limit, we will not attempt to find the explicit Bogoliubov transform, but rather sketch how to relate the corresponding tunneling rates $\Gamma(\omega)$ and 
$\tilde{\Gamma}(\omega)$ with each other.
In fact, a relation between them can be found by using functional calculus methods.
Let us define the Cauchy transform of a function $\Gamma(\omega)$ for $z\in \mathbb{C}$ via
\begin{align}
W(z) \equiv  \int \frac{\Gamma(\omega)}{z + \omega}  \frac{d\omega}{2\pi} \,.
\end{align}
Then, it holds that 
\begin{align}
\lim_{\delta\to 0^+} & W(-\omega+ \ii\delta) \nn 
& = \lim_{\delta\to 0^+}\int \frac{\Gamma(\omega')}{\omega'-\omega+\ii\delta} \frac{d\omega'}{2\pi}\nn
&= \lim_{\delta\to 0^+}\int  \frac{\Gamma(\omega')(\omega'-\omega-\ii\delta)}{(\omega'-\omega)^2+\delta^2} \frac{d\omega'}{2\pi} \nn
&= {\cal P} \int \frac{\Gamma(\omega')}{\omega'-\omega} \frac{d\omega'}{2\pi} - \frac{\ii}{2} \Gamma(\omega)\,.
\end{align}
Hence the imaginary part of the Cauchy transform \mbox{$W(-\omega+\ii\delta)$} can be used to recover the original function $\Gamma(\omega)$.

Now, for a mapping between Eq.~(\ref{EQ:ham1}) and Eq.~(\ref{EQ:ham2}), we can -- in the limit of a continuum of reservoir modes -- 
use  Eq.~(C6) in Ref.~\cite{nazir2018a} to relate the Cauchy transforms of original and mapped (marked by a $\,\tilde{\phantom{c}}\,$ symbol) models via
\begin{align}
\tilde{W}(z) = -\frac{\lambda^2}{z+\epsilon-W(z)}\,.
\end{align}
Evaluating this formula at $z=-\omega+\ii\delta$, we obtain a mapping relation between the tunneling rates
\begin{align}\label{EQ:mapping0}
\tilde{\Gamma} (\omega) =
\frac{\lambda^2 \Gamma(\omega)}
{\left(\omega-\epsilon-{\cal P}\int \frac{\Gamma(\omega')}{\omega-\omega'} \frac{d\omega'}{2\pi}\right)^2
+\left(\frac{\Gamma(\omega)}{2}\right)^2}\,.
\end{align}
Here, one observes that the additional structure in the transformed spectral density $\tilde{\Gamma}(\omega)$ is determined by the coupling strength $\lambda$ and the on-site energy $\epsilon$.
Further, if $H_{\rm rest}$ in~(\ref{EQ:ham1}) consists only of on-site energies and next-neighbor tunnel terms (i.e., non-interacting electrons), the mapping can be performed recursively by redefining the boundary between system and reservoir after every step.
Formally, this corresponds to 
replacing $\lambda \to \lambda_n$, $\epsilon \to \epsilon_n$,
$\tilde{\Gamma}(\omega)\to\Gamma^{(n-1)}(\omega)$, and $\Gamma(\omega) \to \Gamma^{(n)}(\omega)$
(compare also Fig.~\ref{FIG:mapping_sketch}), such that 
the mapping relation reads
\begin{align}\label{EQ:mapping}
\Gamma&^{(n-1)} (\omega) = \nn
& \frac{\lambda_{n}^2 \Gamma^{(n)}(\omega)}
{\left(\omega-\epsilon_{n}-{\cal P}\int \frac{\Gamma^{(n)}(\omega')}{\omega-\omega'} \frac{d\omega'}{2\pi}\right)^2
+\left(\frac{\Gamma^{(n)}(\omega)}{2}\right)^2}\,.
\end{align}
This reverse iteration relation is our first result (where, to remain consistent with the forward mapping~\cite{martinazzo2011a,strasberg2016a,nazir2018a}, 
we have chosen to decrease the index on the l.h.s.).
Instead of calculating the explicit Bogoliubov transform of the mapping -- which would become infeasible for infinite reservoirs, the transformed spectral densities can be obtained by functional calculus or numerical integration.
To arrive at this formula, we have used the very same techniques that are applied in the derivation of the forward mapping, just with the different objective of transferring system degrees of freedom to the reservoirs.

We would like to summarize a few properties of the reverse reaction coordinate mapping:
As with the forward reaction-coordinate mapping, we see that a rigid cutoff of $\Gamma^{(N)}(\omega)$ is also preserved under the reverse mapping.
Likewise, a semicircle spectral density $\Gamma(\omega) = \delta \sqrt{1-(\omega-\epsilon)^2/\delta^2}$ is invariant under the mapping procedure, 
provided that all chain energies and tunneling amplitudes are identical $\epsilon_n=\epsilon$ and $\lambda_n = \delta/2$.
We can further check for particular examples e.g. in Table II of Ref.~\cite{nazir2018a} that this recursion formula reverses the original reaction coordinate mapping.
In particular, when applied to flat reservoir spectral densities, the first application of the mapping just yields a Lorentzian as used in many non-Markovian models~\cite{kleinekathoefer2004a}.
For a large class of spectral densities one can show that with each application of the reverse mapping, at most a single additional peak will be added in the transformed spectral density, see App.~\ref{APP:pole_structure}.

%%%%%%%%%%%%%%%%%%%%%%%%%%%%%%%%%%%%%%%%%%%%%%%%%%%%%%%%%%%%%%%%%%%%%%%%%%%%%%%%%%%%%%%%%%%%%%%%%%%

\subsection{Chain models}

We are interested in chain-type systems connected to two reservoirs, which are held at local thermal equilibrium states.
A non-equilibrium scenario can be generated by applying e.g. different temperatures and/or chemical potentials to the leads, which 
would enter the Fermi functions in Eq.~(\ref{EQ:cgf}).
In total, the Hamiltonian thus reads
$H = H_L + H_{c,L} + H_c + H_{c,R} + H_R$,
where at the ends we have the reservoir parts that can be modeled by non-interacting spinless (or spin-polarized) fermions 
$H_\alpha = \sum_k \epsilon_{k\alpha} c_{k\alpha}^\dagger c_{k\alpha}$
with on-site energies $\epsilon_{k\alpha}$ and $\alpha\in\{L,R\}$.
The central part is a tight-binding chain of length $N$
\begin{align}
H_c = \sum_{i=1}^N \varepsilon_i c_i^\dagger c_i + \sum_{i=1}^{N-1} t_i \left(c_i^\dagger c_{i+1} + c_{i+1}^\dagger c_i\right)\,,
\end{align}
described by on-site energies $\varepsilon_i$ and next-neighbor tunneling amplitudes $t_i>0$ (possible phases of the tunneling
amplitudes can in 1d models be absorbed by unitary transformations of the annihilation and creation operators).
At its ends, the chain is tunnel-coupled to the reservoirs
\begin{align}
H_{c,L} &= \sum_{k} \left(t_{kL} c_{kL}^\dagger c_1 + t_{kL}^* c_1^\dagger c_{kL}\right)\,,\nn
H_{c,R} &= \sum_{k} \left(t_{kR} c_N^\dagger c_{kR} + t_{kR}^* c_{kR}^\dagger c_N\right)
\end{align}
via the tunneling amplitudes $t_{k\alpha}$.
For each reservoir, we define the initial spectral coupling densities in the usual way
$\Gamma_\alpha^{(N)}(\omega) = 2\pi \sum_k \abs{t_{k\alpha}}^2 \delta(\omega-\epsilon_{k\alpha})$
but the index $(N)$ indicates that we take these coupling densities as the first starting point of our iteration, and that we are going
to apply at most $(N-1)$ mappings for a system of $N$ quantum dots.
Since the Bogoliubov transform only involves the boundary mode and the reservoir modes, independent mappings can be individually performed on the reservoirs until the central chain is reduced to a single dot -- which can be arbitrarily chosen along the chain.
For a chain of five quantum dots, we could e.g. independently map the leftmost two dots to the left reservoir and the rightmost two dots into the
right reservoir, leaving only the central dot in a structured single-electron transistor setup.
Alternatively, we could perform four consecutive mappings of the leftmost four dots into the left reservoir, leaving only the rightmost dot as the system.
The transmission of both procedures is the same: The additional few degrees of freedom in the reservoir
cannot change the stationary current, although one may of course expect deviations in the temporal
dynamics.
To perform the mappings, we just mention that one has to identify for mappings of the left reservoir
$\lambda_n = t_{N-n+1}$ and $\epsilon_n = \varepsilon_{N-n+1}$
and for mappings of the right reservoir 
$\lambda_n = t_{n-1}$ and $\epsilon_n = \varepsilon_n$.

For a chain of $N$ sites that is initially coupled to two reservoirs with flat spectral densities, 
the first mapping will just yield Lorentzian-shaped spectral densities, to which the arguments of
App.~\ref{APP:pole_structure} apply.
Therefore, having transferred $N_L$ and $N_R$ sites (with $N_L+N_R=N-1$) to the left and right reservoirs, respectively, the resulting spectral densities for the single leftover dot will have at most 
$N_L$ and $N_R$ peaks, corresponding to $2 N_L$ and $2 N_R$ poles in the complex plane, as pointed out in App.~\ref{APP:pole_structure}.
Now, Eq.~(\ref{EQ:transmission}) and Eq.~(\ref{EQ:mapping0}) are formally very similar, such that it does not come as a surprise, that, with using the same arguments, the transmission of a chain of $N$ dots coupled at its ends to reservoirs with flat spectral densities can have at most $N$ peaks, as is observable in multiple examples below.

%%%%%%%%%%%%%%%%%%%%%%%%%%%%%%%%%%%%%%%%%%%%%%%%%%%%%%%%%%%%%%%%%%%%%%%%%%%%%%%%%%%%%%%%%%%%%%%%%%%
%%%%%%%%%%%%%%%%%%%%%%%%%%%%%%%%%%%%%%%%%%%%%%%%%%%%%%%%%%%%%%%%%%%%%%%%%%%%%%%%%%%%%%%%%%%%%%%%%%%
%%%%%%%%%%%%%%%%%%%%%%%%%%%%%%%%%%%%%%%%%%%%%%%%%%%%%%%%%%%%%%%%%%%%%%%%%%%%%%%%%%%%%%%%%%%%%%%%%%%
%%%%%%%%%%%%%%%%%%%%%%%%%%%%%%%%%%%%%%%%%%%%%%%%%%%%%%%%%%%%%%%%%%%%%%%%%%%%%%%%%%%%%%%%%%%%%%%%%%%
\section{Examples}\label{SEC:examples}

In this section, we discuss applications of the mapping relation~(\ref{EQ:mapping}).
We follow the convention that $\Gamma^{(N)}_\alpha(\omega)$ labels the initial spectral density of reservoir $\alpha$ that is valid for a system size
of $N$ quantum dots, such that $\Gamma^{(n)}_\alpha(\omega)$ will only be evaluated down to $n=1$ in the most extreme cases.
We will also transfer the notation to the final level renormalization needed in Eq.~(\ref{EQ:transmission})
\begin{align}
\Sigma^{(n_L n_R)}(\omega) &= \Sigma_L^{(n_L)}(\omega) + \Sigma_R^{(n_R)}(\omega)\,,\nn
\Sigma_\alpha^{(n_\alpha)}(\omega) &= \frac{1}{2\pi} {\cal P} \int \frac{\Gamma_\alpha^{(n_\alpha)}(\omega')}{\omega-\omega'}d\omega'\,.
\end{align}
We will always map the chain configurations to a single quantum dot, implying that we consider mappings $\Gamma_\alpha^{(n_\alpha)}(\omega)$ 
with the additional constraint $n_L+n_R=N+1$.

%%%%%%%%%%%%%%%%%%%%%%%%%%%%%%%%%%%%%%%%%%%%%%%%%%%%%%%%%%%%%%%%%%%%%%%%%%%%%%%%%%%%%%%%%%%%%%%%%%%

\subsection{Analytic tripledot transmission}

\begin{figure}\centering
\includegraphics[width=0.3\textwidth,clip=true]{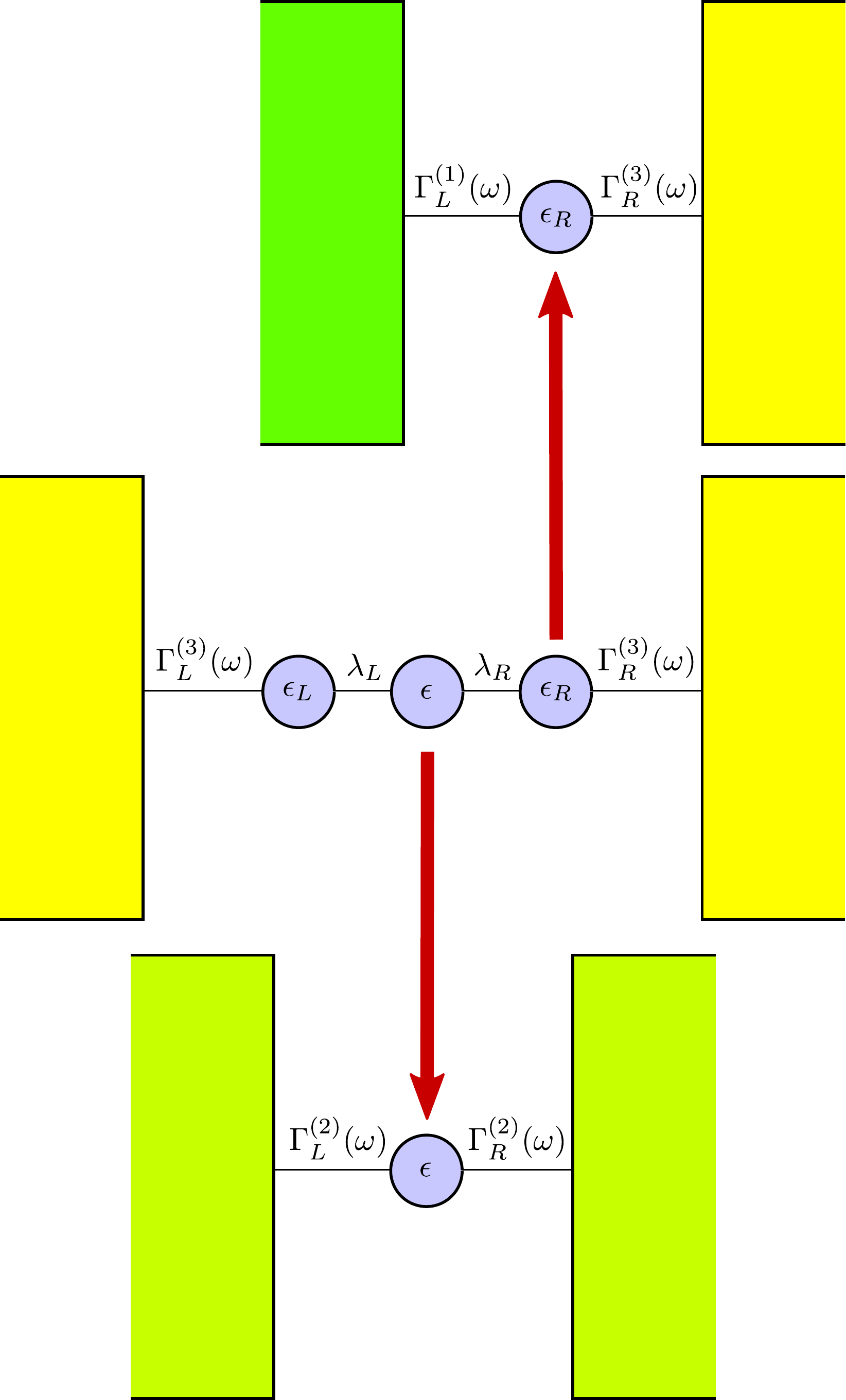} 
\caption{\label{FIG:sketch_tqd}
Illustration of mapping possibilities between a triple quantum dot and the single electron transistor.
Both yield the same transmission.}
\end{figure}

It is well-known that a double quantum dot with flat spectral densities can be mapped to a single quantum dot where one of the
leads has a Lorentzian-shaped spectral density~\cite{zedler2009a}.
We do now illustrate that a triple quantum dot
\begin{align}
H_S = &\epsilon_L d_L^\dagger d_L + \epsilon d^\dagger d + \epsilon_R d_R^\dagger d_R\nn
&+ \lambda_L \left(d_L^\dagger d + d^\dagger d_L\right) + \lambda_R \left(d^\dagger d_R + d_R^\dagger d\right)\,,
\end{align}
serially connecting two reservoirs with flat spectral densities can be mapped to a single quantum dot coupled to two reservoirs, either with
both having Lorentzian spectral densities or one with a flat spectral density and the other one with a more structured spectral density, see 
Fig.~\ref{FIG:sketch_tqd}.

We start with a flat spectral density for both reservoirs
\begin{align}
\Gamma_\alpha^{(3)}(\omega) = \Gamma_\alpha\,,
\end{align}
and perform a single mapping for each reservoir.
Inserting this in the mapping~(\ref{EQ:mapping}), the transformed mapping becomes
\begin{align}
\Gamma_\alpha^{(2)}(\omega) = \frac{\lambda_\alpha^2 \Gamma_\alpha}{(\omega-\epsilon_\alpha)^2 + \left(\frac{\Gamma_\alpha}{2}\right)^2}\,.
\end{align}
The level renormalization becomes $\Sigma^{22}(\omega)=\Sigma^{2}_L(\omega)+\Sigma^{2}_R(\omega)$ with
\begin{align}
\Sigma^{2}_\alpha(\omega) = \frac{\lambda_\alpha^2 (\omega-\epsilon_\alpha)}{(\omega-\epsilon_\alpha)^2 + \left(\frac{\Gamma_\alpha}{2}\right)^2} \,.
\end{align}
We can insert these two equations into the transmission formula~(\ref{EQ:transmission}), obtaining the transmission of a triple quantum dot 
analytically (not shown), which enables us to perform optimizations.

Supposing that one has control only over the $\epsilon_\alpha$ and $\lambda_\alpha$ but not the external coupling strengths $\Gamma_\alpha$, one choice
that satisfies the constraints $\Sigma^{(22)}(\epsilon)=0$ and $\Gamma_L^{(2)}(\epsilon)=\Gamma_R^{(2)}(\epsilon)$ would be
$\epsilon_L=\epsilon_R=\epsilon$ and $\lambda_R=\sqrt{\Gamma_R / \Gamma_L} \lambda_L$. 
With this choice, the transmission would have a peak at $\omega=\epsilon$ with maximum value $T(\epsilon)=1$ even for highly asymmetric tunneling rates, i.e., by optimizing the internal chain parameters one can correct for an unfavorably asymmetric external coupling.
Another choice -- if the external tunneling rates can also be controlled -- would be to choose $\Gamma_L=\Gamma_R=\Gamma$ and $\epsilon_L=\epsilon_R=\epsilon$ as well as $\lambda_L=\lambda_R=\lambda\to\infty$, which generates a singly-peaked transmission 
$T(\omega) \to \frac{\Gamma^2}{\Gamma^2 + 4(\omega-\epsilon)^2}$
with width $\Gamma$ and $T(\epsilon)=1$.

\begin{figure}
\includegraphics[width=0.45\textwidth,clip=true]{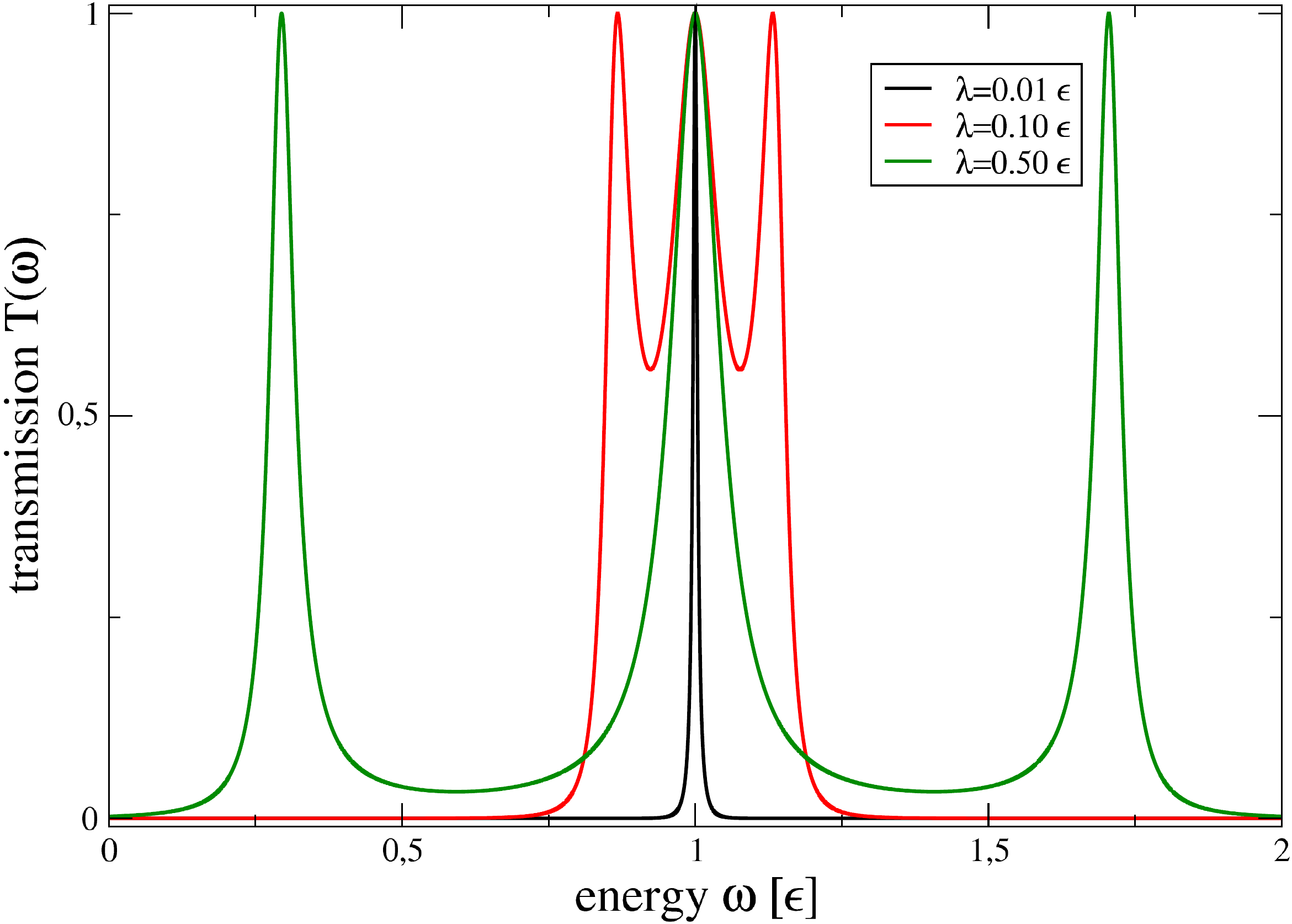}\\
\includegraphics[width=0.45\textwidth,clip=true]{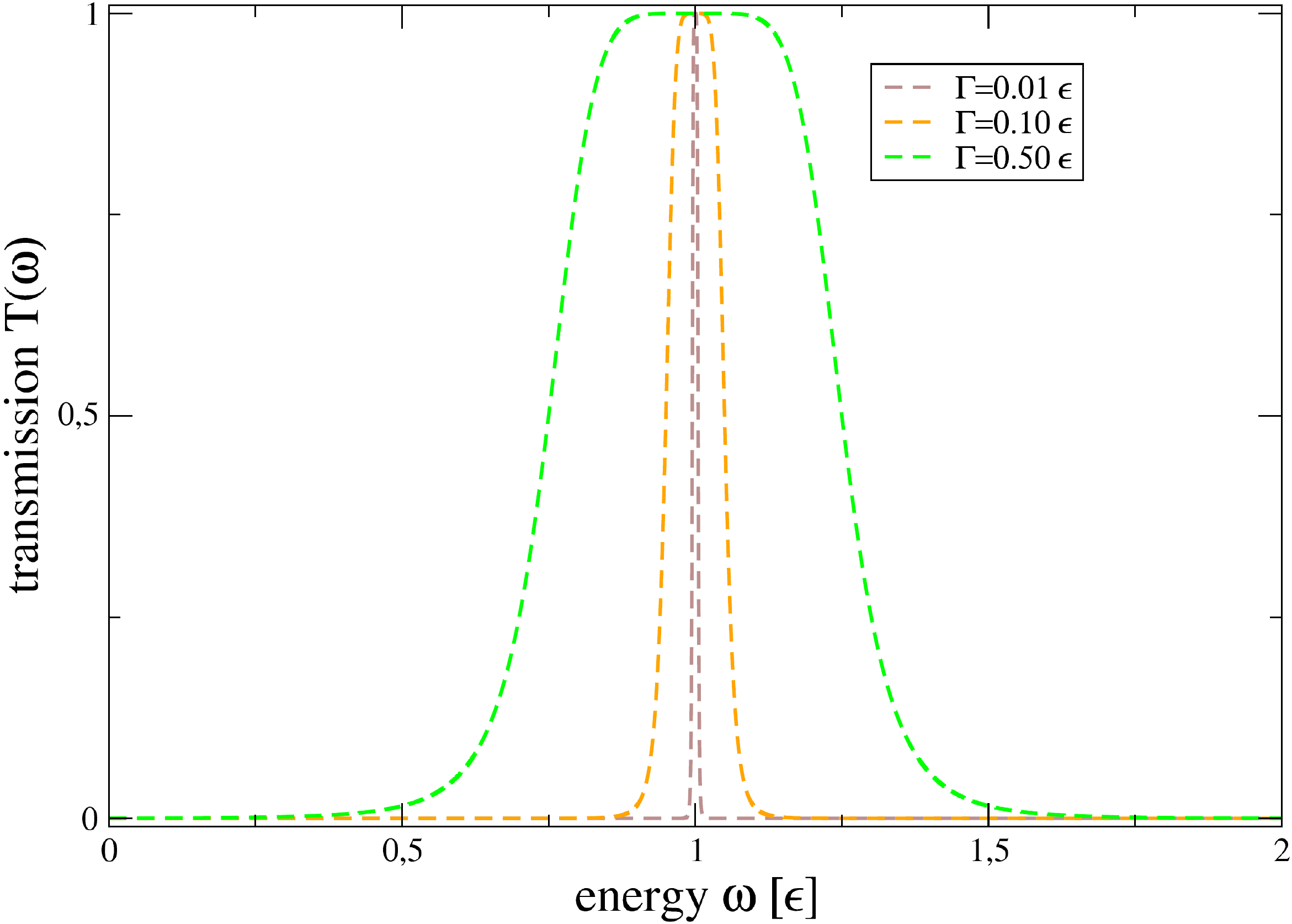}
\caption{\label{FIG:transmission_tqd}
{\bf (top)}
Transmission of a triple quantum dot versus energy for different internal coupling strengths $\lambda_L=\lambda_R=\lambda$ and
moderate external coupling $\Gamma_\alpha=0.1 \epsilon$ (solid curves).
When the internal coupling strength $\lambda$ is increased while keeping the external strength $\Gamma$, the transmission peak splits into three
(solid black, red, and dark green).
{\bf (bottom)}
Same as top right, for different external coupling strengths $\Gamma_L=\Gamma_R=\Gamma$.
When the internal coupling strength is tied to the external one $\lambda = \Gamma/\sqrt{8}$, one may achieve a near box-shaped transmission
(dashed brown, orange, and light green). 
Other parameters: $\epsilon_L=\epsilon_R=\epsilon$ (all curves).
}
\end{figure}

To generate a transmission that is flat around $\epsilon$, one may Taylor-expand the denominator of Eq.~(\ref{EQ:mapping})
and demand that the first orders in $\omega-\epsilon$ vanish, which yields constraints on the tunneling amplitudes.
For example, under the assumption of symmetric external couplings $\Gamma_L=\Gamma_R=\Gamma$, a suitable choice would be $\epsilon_L=\epsilon_R=\epsilon$ and $\lambda_L=\lambda_R=\Gamma/\sqrt{8}$, yielding the transmission
\begin{align}\label{EQ:engtransmission_3}
T(\omega) = \frac{\Gamma^6}{\Gamma^6 + 64 (\omega-\epsilon)^6}\,,
\end{align}
which is flat around $\omega=\epsilon$.
The triple quantum dot transmission is illustrated for different parameters in Fig.~\ref{FIG:transmission_tqd}.

Just for consistency, we sketch that by mapping 
e.g. both the left and the central dot to the left reservoir and leaving the right dot unchanged, we obtain the same transmission.
Then, naturally, the right spectral density will remain unmodified $\Gamma_R(\omega) = \Gamma_R^{(3)}(\omega)$, 
whereas for the left reservoir, the iteration yields
\begin{align}
\Gamma_L^{(2)}(\omega) &= \frac{\lambda_L^2 \Gamma_L}{(\omega-\epsilon_L)^2 + \left(\frac{\Gamma_L}{2}\right)^2}\,,\nn
\Gamma_L^{(1)}(\omega) &= \frac{\Gamma_L \lambda_L^2 \lambda_R^2}
{\gamma_{1,L}(\omega) - 8 (\omega-\epsilon) (\omega-\epsilon_L) \lambda_L^2 + 4 \lambda_L^4}\,,\nn
\gamma_{1,L}(\omega) &\equiv  (\omega-\epsilon)^2\left(\Gamma_L^2 + 4 (\omega-\epsilon_L)^2\right)\,.
\end{align}
Here, we see that $\Gamma_L^{(1)}(\omega)$ will have at most two peaks, compare App.~\ref{APP:pole_structure}.
The level renormalization $\Sigma^{(13)}(\omega)=\Sigma^{(1)}_L(\omega)+\Sigma^{(3)}_R(\omega)$ induced by the right reservoir still vanishes $\Sigma^{(3)}_R(\omega)=0$, 
but the one of the left reservoir corrects for the different tunneling rates
\begin{align}
\Sigma^{(1)}_L(\omega) &= 
\frac{\lambda_R^2 \left(\Gamma_L^2 (\omega-\epsilon )+\gamma_{2,L}(\omega)\right)}
{\Gamma_L^2 (\epsilon -\omega)^2+4 \left(\lambda_L^2+(\epsilon -\omega) (\omega-\epsilon_L)\right)^2}\,,\nn
\gamma_{2,L}(\omega) &\equiv  4 (\epsilon_L-\omega) \left(\lambda_L^2+(\epsilon -\omega) (\omega-\epsilon_L)\right)\,.
\end{align}
With Eq.~(\ref{EQ:transmission}) we can check that the resulting lengthy expression for the transmission is fully identical with the 
one from the previous section (not shown).
In addition, we mention here that the transmission is identical with the one that can be obtained numerically from using non-equilibrium Greens function calculations (compare App.~\ref{APP:greens_function}).

%%%%%%%%%%%%%%%%%%%%%%%%%%%%%%%%%%%%%%%%%%%%%%%%%%%%%%%%%%%%%%%%%%%%%%%%%%%%%%%%%%%%%%%%%%%%%%%%%%%

\subsection{Analytic quintuple dot transmission}

\begin{figure}[ht]\centering
\includegraphics[width=0.4\textwidth,clip=true]{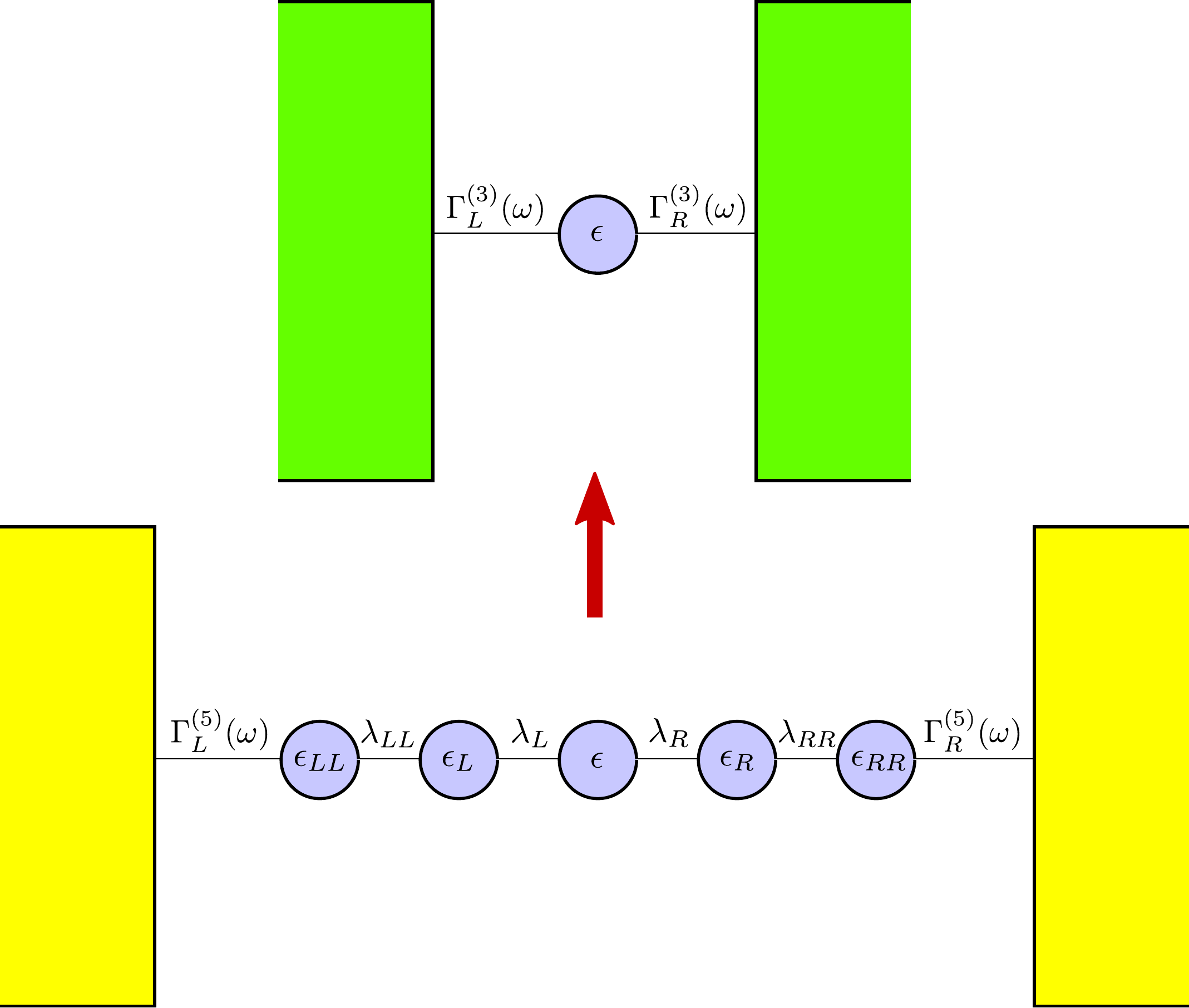} 
\caption{\label{FIG:sketch_pqd}
Illustration of the reverse reaction-coordinate mapping for a quintuple quantum dot.
}
\end{figure}

The same considerations can be applied to five quantum dots that are serially connected
\begin{align}
H_S &=\epsilon_{LL} d_{LL}^\dagger d_{LL} +  \lambda_{LL} \left(d_{LL}^\dagger d_L + d_L^\dagger d_{LL} \right) \nn
&\qquad+\epsilon_L d_L^\dagger d_L +  \lambda_L \left(d_L^\dagger d + d^\dagger d_L\right)  \nn
&\qquad+\epsilon d^\dagger d + \lambda_R \left(d^\dagger d_R + d_R^\dagger d\right) \nn
&\qquad+\epsilon_R d_R^\dagger d_R  + \lambda_{RR} \left(d_R^\dagger d_{RR} + d_{RR}^\dagger d_R\right)\nn
&\qquad+\epsilon_{RR} d_{RR}^\dagger d_{RR}
\,.
\end{align}
Here, we can map the two boundary dots into the reservoirs, which become structured, leaving only the central dot as the system
is illustrated in Fig.~\ref{FIG:sketch_pqd}. 

Starting again from flat spectral densities, the mapping relations yield the tunneling rates
\begin{align}
\Gamma_\alpha^{(5)}(\omega) &= \Gamma_\alpha\,,\nn
\Gamma_\alpha^{(4)}(\omega) &= \frac{\lambda_{\alpha\alpha}^2 \Gamma_\alpha}{(\omega-\epsilon_{\alpha\alpha})^2+\frac{\Gamma_\alpha^2}{4}}\,,\nn
\Gamma_\alpha^{(3)}(\omega) &= \frac{4\Gamma_\alpha \lambda_\alpha^2 \lambda_{\alpha\alpha}^2}{\Gamma_\alpha^2(\omega-\epsilon_\alpha)^2
+4 [\gamma_{3,\alpha}(\omega)]^2}\,,\nn
\gamma_{3,\alpha}(\omega) &\equiv \lambda_{\alpha\alpha}^2 - (\omega-\epsilon_\alpha)(\omega-\epsilon_{\alpha\alpha})\,.
\end{align}
We see that $\Gamma_\alpha^{(3)}(\omega)$ has two more poles than $\Gamma_\alpha^{(4)}(\omega)$, compare
App.~\ref{APP:pole_structure}.
From these we can explicitly calculate the level renormalization $\Sigma^{(33)}(\omega) = \Sigma^{(3)}_L(\omega)+\Sigma^{(3)}_R(\omega)$ where
\begin{align}
\Sigma^{(3)}_\alpha(\omega) &= \frac{\lambda_\alpha^2 \left((\omega-\epsilon_\alpha)\left(\Gamma_\alpha^2+4(\omega-\epsilon_{\alpha\alpha})^2\right)
-\gamma_{4,\alpha}(\omega)\right)}
{\Gamma_\alpha^2 (\omega-\epsilon_\alpha)^2 + 4 [\gamma_{3,\alpha}(\omega)]^2}\,,\nn
\gamma_{4,\alpha}(\omega) &\equiv 4\lambda_{\alpha\alpha}^2 (\omega-\epsilon_{\alpha\alpha})\,.
\end{align}
Inserting this into the single-dot transmission~(\ref{EQ:transmission}), we obtain an analytic expression for the transmission through a chain 
of five quantum dots.

\begin{figure}
\includegraphics[width=0.45\textwidth,clip=true]{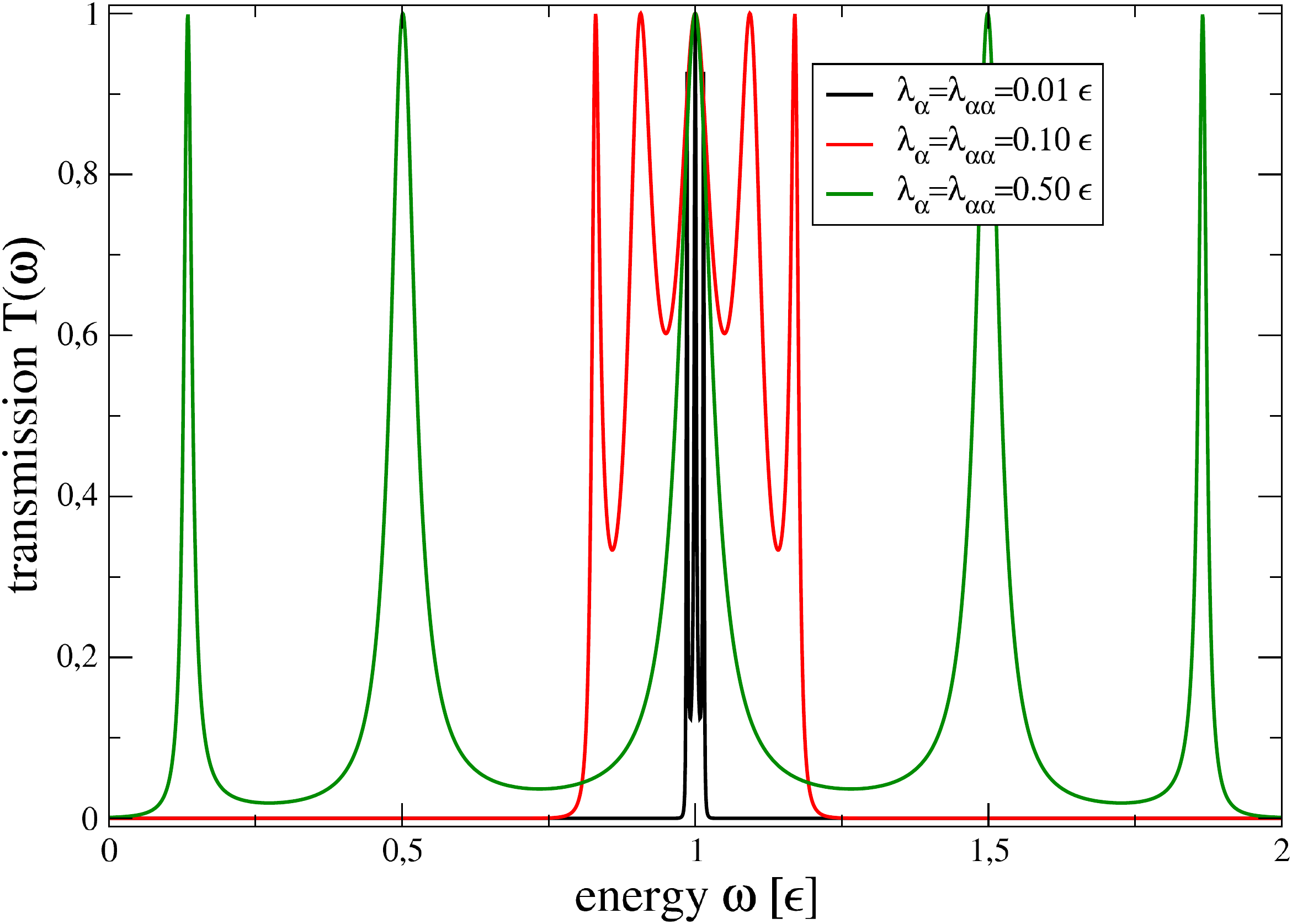} 
\includegraphics[width=0.45\textwidth,clip=true]{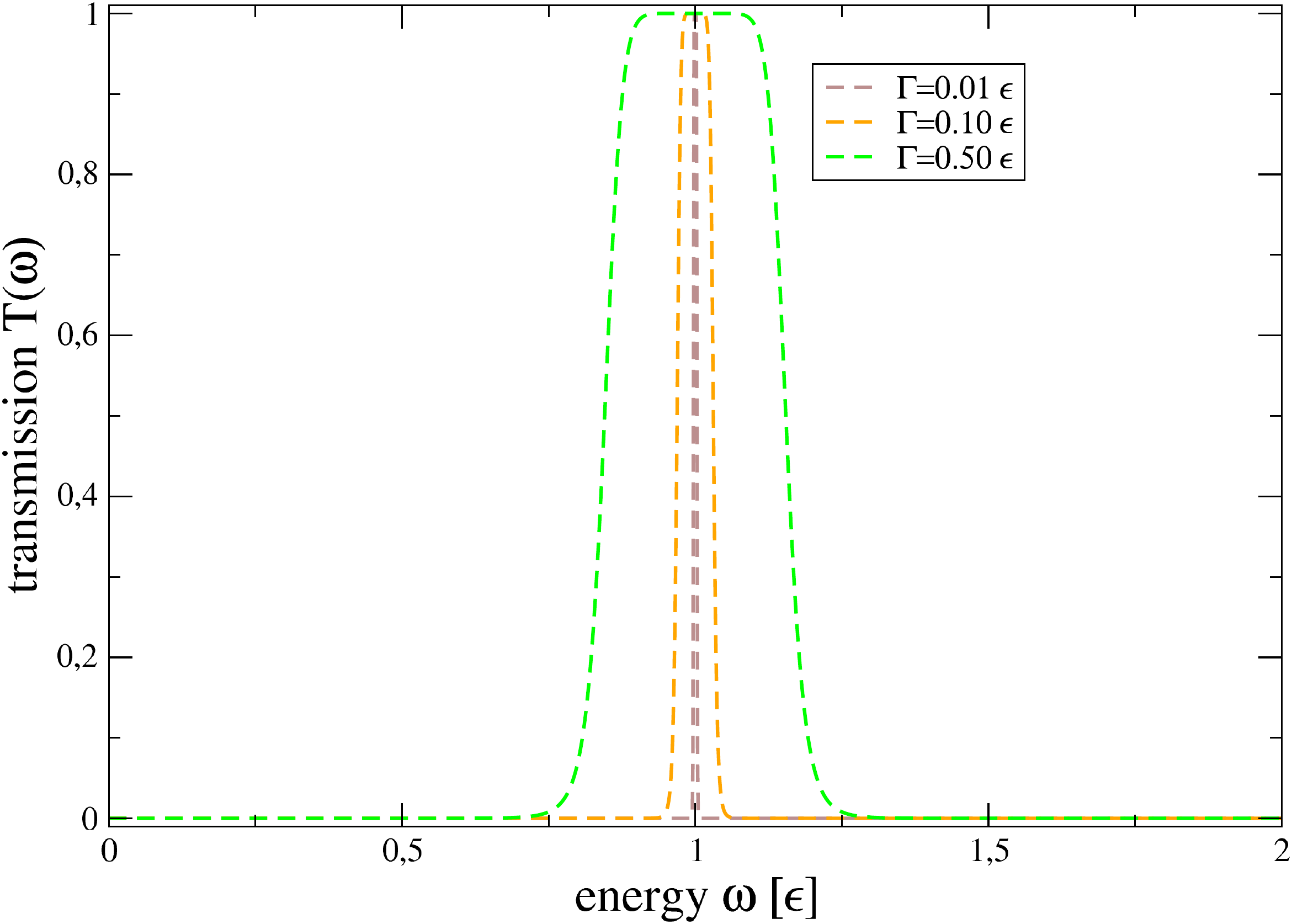}
\caption{\label{FIG:transmission_pqd}
{\bf (top)}
Transmission of a quintuple dot chain versus energy for different internal coupling strengths $\lambda_\alpha$ and 
$\lambda_{\alpha\alpha}$ and external couplings  $\Gamma_L=\Gamma_R= 0.1 \epsilon$ (solid curves).
{\bf (bottom)}
Same as middle, but for different external coupling strengths $\Gamma_L=\Gamma_R=\Gamma$ and internal couplings tied to the external 
ones $\lambda_{\alpha} = \frac{\Gamma}{2} \sqrt{\frac{\sqrt{5}-2}{2}}$ and $\lambda_{\alpha\alpha}=\frac{\sqrt{5}-1}{4} \Gamma$ (dashed curves).
The effects are enhanced in comparison to a triple quantum dot in Fig.~\ref{FIG:transmission_tqd} right panels, and a box-shaped transmission 
is approached for strong system-lead couplings.
Other parameters: $\epsilon_L=\epsilon_R=\epsilon=\epsilon_{LL}=\epsilon_{RR}$ (all curves).
}
\end{figure}

In order to generate a transmission centered around $\epsilon$ with maximum value $T(\epsilon)=1$ and increased broadness, we can  use Taylor expansion
of the denominator in the mapping~(\ref{EQ:mapping}), where the choice  
$\epsilon_\alpha=\epsilon_{\alpha\alpha} = \epsilon$, $\Gamma_\alpha=\Gamma$, $\lambda_\alpha=\frac{\Gamma}{2} \sqrt{\frac{\sqrt{5}-2}{2}}$, and 
$\lambda_{\alpha\alpha} = \frac{\sqrt{5}-1}{4} \Gamma$ yields a transmission that is flat around $\epsilon$
\begin{align}\label{EQ:engtransmission_5}
T(\omega) = \frac{(123-55 \sqrt{5}) \Gamma^{10}}{(123-55 \sqrt{5}) \Gamma^{10} + 2^{11} (\omega-\epsilon)^{10}}\,.
\end{align}
This is already quite close to the previously mentioned box-shaped example.
The transmission is illustrated for different parameters in Fig.~\ref{FIG:transmission_pqd}, and 
we have explicitly checked complete agreement with numerical non-equilibrium Greens function calculations (not shown, compare App.~\ref{APP:greens_function}).

%%%%%%%%%%%%%%%%%%%%%%%%%%%%%%%%%%%%%%%%%%%%%%%%%%%%%%%%%%%%%%%%%%%%%%%%%%%%%%%%%%%%%%%%%%%%%%%%%%%
\subsection{Non-Markovian transport signatures at steady state}

\begin{figure}[ht]
\includegraphics[width=0.45\textwidth,clip=true]{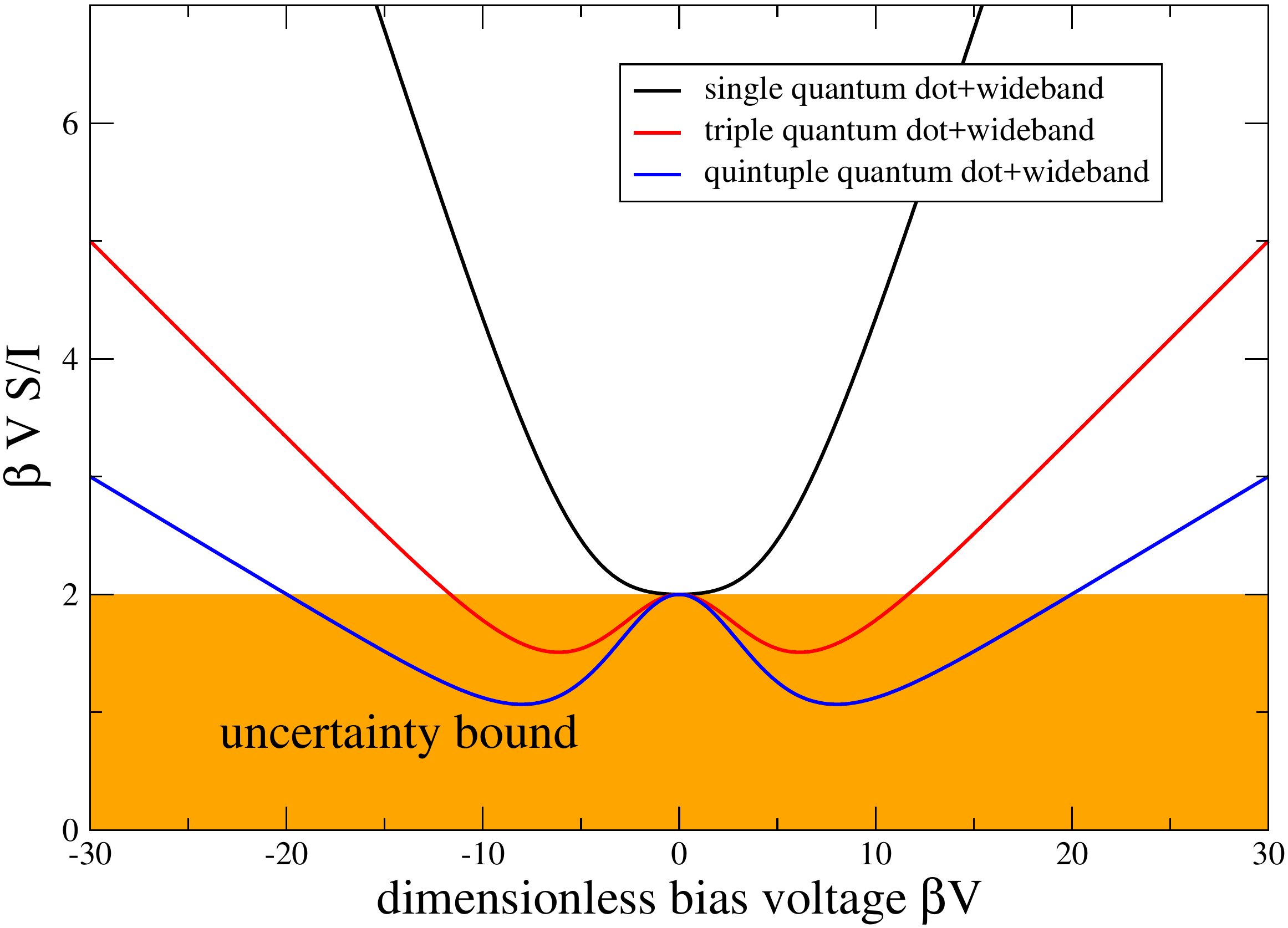} 
\caption{\label{FIG:uncertainty}
Plot of the l.h.s. of inequality~(\ref{EQ:bound}) versus bias voltage.
The violation of the classical Markovian uncertainty relation~(\ref{EQ:bound}) -- marked orange region -- is possible for engineered transmission of the 
triple~(\ref{EQ:engtransmission_3}) and quintuple~(\ref{EQ:engtransmission_5}) quantum dots (red and blue, respectively).
Parameters: $\beta_L=\beta_R=\beta$, $\mu_L=-\mu_R=V/2$, $\beta\Gamma=1.0$, $\epsilon=0$.
}
\end{figure}

The engineered transmissions derived for triple and quintuple quantum dots indeed violate the thermodynamic uncertainty bound that classical systems (described by
Markovian transition rates obeying detailed balance) must obey~\cite{barato2015a,pietzonka2016a,gingrich2016a,horowitz2017a,pietzonka2018a}.
It relates the entropy production rate with the ratio of noise and squared current.
In particular, at equal temperatures $\beta_L=\beta_R=\beta$ and with bias voltage $V=\mu_L-\mu_R$, 
the entropy production rate becomes $\sigma = \beta I V$, such that 
the thermodynamic uncertainty relation reads
\begin{align}\label{EQ:bound}
\beta V \frac{S}{I} = \beta \abs{V} F \ge 2\,,
\end{align}
where $S$ and $I$ are noise and current, respectively, and $F=S/\abs{I}$ is the Fano factor.
That means, a violation of this bound is a hallmark of truly non-classical behaviour~\cite{agarwalla2018a}.
In particular, it would be interesting to investigate whether heat engines built with such 
devices can overcome standard bounds~\cite{pietzonka2018a}.
We can evaluate the uncertainty relation by computing Eqns.~(\ref{EQ:current_noise}) with the transmission of a single quantum dot~(\ref{EQ:transmission})
and flat tunneling rates and compare with the engineered transmissions of triple~(\ref{EQ:engtransmission_3}) and quintuple quantum dots~(\ref{EQ:engtransmission_5}).
Whereas -- as noted in~\cite{agarwalla2018a} -- a single quantum dot in the wide-band limit does not violate the bound~(\ref{EQ:bound}), engineered triple- and quintuple quantum dots
may do so in a large bias voltage window, see Fig.~\ref{FIG:uncertainty}.
We do actually observe this behaviour also for significantly smaller coupling strengths $\beta \Gamma \ll 1$, demonstrating that non-Markovian behaviour can also occur in the weak-coupling regime (not shown).
Thus, while for weak chain-reservoir couplings, a time-local master equation description of the chain exists, it is still non-Markovian in the sense that the secular approximation, which eventually is required to generate Lindblad-type Markovian master equations, is not applicable.
Indeed, to obtain the transmissions~(\ref{EQ:engtransmission_3}) and~(\ref{EQ:engtransmission_5}), the level spacings of the tripledot or quintupledot chains need to be comparable to the chain lead couplings, invalidating the secular approximation.

%%%%%%%%%%%%%%%%%%%%%%%%%%%%%%%%%%%%%%%%%%%%%%%%%%%%%%%%%%%%%%%%%%%%%%%%%%%%%%%%%%%%%%%%%%%%%%%%%%%
\subsection{SSH chain}

In this section, we will demonstrate that the mapping can also be performed numerically.
As an example, we consider the celebrated Su-Schrieffer-Heeger (SSH) model~\cite{su1979a}, which corresponds to a chain of dimers.
Within the tight-binding representation, this can be written by alternating tunneling amplitudes $\tau_i$, e.g. for $M$ dimers corresponding
to $N=2M$ quantum dots
\begin{align}\label{EQ:hamssh}
H_S = \epsilon \sum_{i=1}^{N} & d_i^\dagger d_i  + \tau_1 \sum_{i=1}^{M} \left[d_{2i-1}^\dagger d_{2i} + d_{2i}^\dagger d_{2i-1}\right]\nn
&+ \tau_2 \sum_{i=1}^M \left[d_{2i}^\dagger d_{2i+1} + d_{2i+1}^\dagger d_{2i}\right]\,.
\end{align}
In the extreme case $\tau_1=0$, one finds two isolated monomers at the chain ends, whereas when $\tau_2=0$, one has a chain of decoupled
dimers.
In the infinite-size limit $N\to\infty$, the model exhibits a topological phase transition at $\tau_1=\tau_2$ between a topological phase ($\tau_1 < \tau_2$) supporting two edge states and a normal phase ($\tau_1>\tau_2$) without edge modes~\cite{asboth2016}.
We open the chain by connecting it to two fermionic reservoirs at its ends.
Thus, we assume that the initial tunneling rates are those resulting from semi-infinite
homogeneous tight-binding chains given by
\begin{align}\label{EQ:specdens_semicircle}
\Gamma_\alpha^{(N)}(\omega) = \frac{\tau_\alpha^2}{\tau_0^2} \sqrt{4 \tau_0^2 - \omega^2}
\end{align}
for $\omega^2\le 4 \tau_0^2$, 
where $\tau_0$ describes the width of the reservoir spectral densities and $\tau_L$ ($\tau_R$) the tunneling amplitude between 
the first (last) site of the SSH chain and the left (right) reservoir.
We can obtain the transmission for the SSH model with non-equilibrium Greens function techniques along the lines of Ref.~\cite{boehling2018a}, see App.~\ref{APP:greens_function}, and
use it to benchmark the transmission obtained by the successive mappings of the SSH chain to a single quantum dot. 

\begin{figure}[ht]
\includegraphics[width=0.45\textwidth,clip=true]{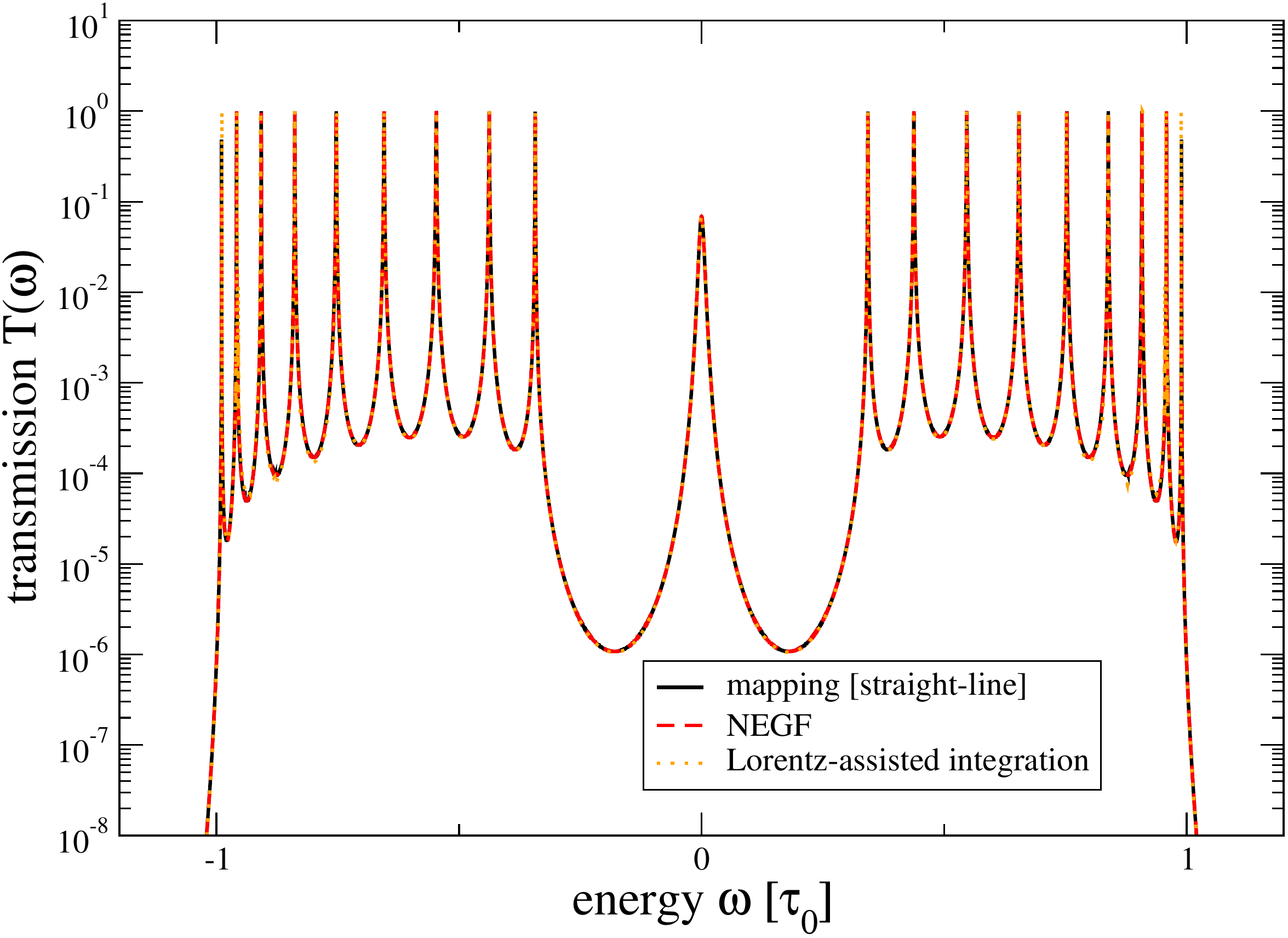}
\caption{\label{FIG:transmissionssh}
Comparison of transmissions of an SSH model~(\ref{EQ:hamssh}), obtained from non-equilibrium Greens function techniques (dashed red, 22000 data points) 
and transmission obtained from the mapping, either with straight-line
approximations (solid black, 40000 data points) or Lorentz-assisted integration (dotted orange, 15000 data points) 
as detailed in App.~\ref{APP:pval1} and App.~\ref{APP:pval2}.
All methods require for this model a fine discretization width in order not to miss particularly sharp peaks.
Other parameters $\epsilon=0.0 \tau_0$, $\tau_1 = 0.35 \tau_0$, $\tau_2 = 0.65 \tau_0$, $\tau_L=\tau_R=0.1 \tau_0$, $N=20$. 
}
\end{figure}

To compute the mapping relation~(\ref{EQ:mapping}) and the level renormalization~(\ref{EQ:level_renormalization}) numerically, 
we employ a straight-line approximation of the tunneling rates, which we outline in Appendix~\ref{APP:pval1}.
In particular since for the chosen parameters the tunneling rates and transmission are composed of multiple very thin peaks, this is 
challenging as the discretization grid has to be chosen very fine or even adaptive to the functional shape:
Since the principal value integral has to be performed for each data point, the total complexity of one mapping iteration scales as
$\ord\{N_{\rm pt}^2\}$ for $N_{\rm pt}$ data points.
In App.~\ref{APP:pval2}, we also outline that by fitting the peaks with Lorentzians, a part of the mapping may be performed in an analytic fashion, 
which allows one to reduce the grid resolution.
The resulting transmission -- computed by performing for a chain of $N=20$ dots ($M=10$ dimers) agrees quite well with the full Greens function
calculation, see Fig.~\ref{FIG:transmissionssh}.
In contrast to the Greens function calculation -- which requires knowledge of the free Greens function (analytically known only for semicircle initial spectral densities) -- the numerical mapping procedure can be applied to arbitrary initial spectral densities with a hard cutoff.

Even stronger, a significant numerical advantage of the method becomes visible via semi-analytic calculation of the transformed spectral densities.
This is possible when e.g. the initial spectral densities are just flat, such that the first mapping already yields a Lorentzian.
Technically, it only requires to find the poles of the mapped spectral density in the complex plane numerically, which can be done iteratively
by a Weierstra{\ss}-Durand-Kerner algorithm
from the poles of the original spectral density. 
We detail this in App.~\ref{APP:pval3}, where we also demonstrate it for the SSH model with initially flat spectral densities.
The associated numerical effort is negligible in comparison to the other approaches, such that for such initial spectral densities, we 
consider this method as the most efficient one.

Beyond benchmarking, we can learn about the SSH model physics by considering the transmission peaks in Fig.~\ref{FIG:transmissionssh}.
First of all, the number of peaks in Fig. \ref{FIG:transmissionssh} agrees well with the arguments in App. \ref{APP:pole_structure}:
We see 19 peaks for a model with 20 sites. 
Therefore, two peaks must have merged, signaling that the system is in the topological phase.
When the current is plotted versus bias voltage, one can from Eq.~(\ref{EQ:current_noise}) conclude that it will -- at sufficiently low temperatures and symmetric chemical potentials $\mu_L=+V/2=-\mu_R$ -- exhibit steps at the peaks of the transmission.
The transmission peak in the center corresponds to the two near-degenerate edge states, it is absent in the normal phase of the model~\cite{boehling2018a}.
The other transmission peaks correspond to the bulk states, and during the finite-size precursor of the topological phase transition, the central transmission peak splits into two.
Thus, the current enables the detection of the topological phase transition point.

%%%%%%%%%%%%%%%%%%%%%%%%%%%%%%%%%%%%%%%%%%%%%%%%%%%%%%%%%%%%%%%%%%%%%%%%%%%%%%%%%%%%%%%%%%%%%%%%%%%
%%%%%%%%%%%%%%%%%%%%%%%%%%%%%%%%%%%%%%%%%%%%%%%%%%%%%%%%%%%%%%%%%%%%%%%%%%%%%%%%%%%%%%%%%%%%%%%%%%%
%%%%%%%%%%%%%%%%%%%%%%%%%%%%%%%%%%%%%%%%%%%%%%%%%%%%%%%%%%%%%%%%%%%%%%%%%%%%%%%%%%%%%%%%%%%%%%%%%%%
%%%%%%%%%%%%%%%%%%%%%%%%%%%%%%%%%%%%%%%%%%%%%%%%%%%%%%%%%%%%%%%%%%%%%%%%%%%%%%%%%%%%%%%%%%%%%%%%%%%
\section{Summary and Outlook}\label{SEC:conclusions}

We have discussed and applied the reverse mapping formula for non-interacting electronic dot chains.
The mapping reverses the original fermionic mapping relation but effectively transfers one degree of freedom (one quantum dot)
from the system into the reservoir.
Thereby, long chain models can be mapped to a single quantum dot that is connected to two structured fermionic reservoirs, with
the tunneling rates obtained from the recursive mapping procedure.
Since the transmission of the single quantum dot is well known, one thereby has an approach to determine the transmission
of chain models from recursive mappings.
Conceptionally, this is very similar to other recursive mapping approaches known for the Greens function itself~\cite{lewenkopf2013a,bonardi2015a}.
However, as the method does not require knowledge of any free Greens function, it can in principle be applied to 
arbitrarily shaped external spectral densities.
For short chains and/or particularly simple initial spectral densities, these mappings can be performed analytically, allowing
for the analytic optimization of the transmission with respect to some desired objective.
Numerically, the mapping can also be performed,
which we demonstrated for the Su-Schrieffer-Heeger chain.
For an initial semicircle spectral density, the numerical calculation of the mapping requires a sufficiently dense sampling, leaving at the
end no significant numerical advantage as compared to the evaluation of the transmission via direct Greens function calculations.
However, for initially flat spectral densities, the mapping can be performed semi-analytically, allowing for an extremely fast 
and accurate calculation of the mapping.
In addition, we would like to stress that the method can also be applied to calculate stationary dot occupations analytically, as these are well-known for a single quantum dot~\cite{haug2008} (and can also be computed for structured tunneling rates~\cite{schaller2009a}).
Other applications of the reverse reaction coordinate mapping are conceivable. 
For example, given an initial chain with spectral densities that correspond to the strong chain-reservoir coupling limit and thereby
inhibit a master equation treatment, the mapping relation can be employed, yielding a shorter 
chain for which a master equation treatment may be applicable, 
allowing to investigate interesting parameter regimes.
In addition, we like to stress that the method is not fundamentally limited to non-interacting electronic transport.
As long as the interacting parts of the system are not involved by the mapping transformations, one may perform them as described.
If for example the transmission in the Meir-Wingreen formula~\cite{meir1992a} of an interacting part of the chain is known, one could
reduce longer chains to the interesting part by successively applying the reverse mapping transformation, significantly reducing the
degrees of freedom that need explicit treatment.
Finally, we think that it could be interesting to find recursion formulas also for two- or higherdimensional systems, which in principle should be possible.

%%%%%%%%%%%%%%%%%%%%%%%%%%%%%%%%%%%%%%%%%%%%%%%%%%%%%%%%%%%%%%%%%%%%%%%%%%%%%%%%%%%%%%%%%%%%%%%%%%%
%%%%%%%%%%%%%%%%%%%%%%%%%%%%%%%%%%%%%%%%%%%%%%%%%%%%%%%%%%%%%%%%%%%%%%%%%%%%%%%%%%%%%%%%%%%%%%%%%%%
%%%%%%%%%%%%%%%%%%%%%%%%%%%%%%%%%%%%%%%%%%%%%%%%%%%%%%%%%%%%%%%%%%%%%%%%%%%%%%%%%%%%%%%%%%%%%%%%%%%
%%%%%%%%%%%%%%%%%%%%%%%%%%%%%%%%%%%%%%%%%%%%%%%%%%%%%%%%%%%%%%%%%%%%%%%%%%%%%%%%%%%%%%%%%%%%%%%%%%%
\section{Acknowledgment}

The authors gratefully acknowledge discussions with S. B\"ohling, L. Pachon, B. K. Agarwalla, P. Strasberg, 
and financial support by the DGF (SFB 910, GRK 1558, BR 1528/9-1).

%%%%%%%%%%%%%%%%%%%%%%%%%%%%%%%%%%%%%%%%%%%%%%%%%%%%%%%%%%%%%%%%%%%%%%%%%%%%%%%%%%%%%%%%%%%%%%%%%%%
%%%%%%%%%%%%%%%%%%%%%%%%%%%%%%%%%%%%%%%%%%%%%%%%%%%%%%%%%%%%%%%%%%%%%%%%%%%%%%%%%%%%%%%%%%%%%%%%%%%
%%%%%%%%%%%%%%%%%%%%%%%%%%%%%%%%%%%%%%%%%%%%%%%%%%%%%%%%%%%%%%%%%%%%%%%%%%%%%%%%%%%%%%%%%%%%%%%%%%%
%%%%%%%%%%%%%%%%%%%%%%%%%%%%%%%%%%%%%%%%%%%%%%%%%%%%%%%%%%%%%%%%%%%%%%%%%%%%%%%%%%%%%%%%%%%%%%%%%%%

\bibliographystyle{unsrt}
\bibliography{references}

\appendix

%%%%%%%%%%%%%%%%%%%%%%%%%%%%%%%%%%%%%%%%%%%%%%%%%%%%%%%%%%%%%%%%%%%%%%%%%%%%%%%%%%%%%%%%%%%%%%%%%%%
%%%%%%%%%%%%%%%%%%%%%%%%%%%%%%%%%%%%%%%%%%%%%%%%%%%%%%%%%%%%%%%%%%%%%%%%%%%%%%%%%%%%%%%%%%%%%%%%%%%
%%%%%%%%%%%%%%%%%%%%%%%%%%%%%%%%%%%%%%%%%%%%%%%%%%%%%%%%%%%%%%%%%%%%%%%%%%%%%%%%%%%%%%%%%%%%%%%%%%%
%%%%%%%%%%%%%%%%%%%%%%%%%%%%%%%%%%%%%%%%%%%%%%%%%%%%%%%%%%%%%%%%%%%%%%%%%%%%%%%%%%%%%%%%%%%%%%%%%%%

\section{Pole structure of the mapping}\label{APP:pole_structure}

To enable an analytic calculation of the principal value integrals in the reverse reaction-coordinate mapping, we do now assume a particular but still fairly general shape of the spectral density, namely to be of the form
\begin{align}\label{EQ:spectral_density}
\Gamma(\omega) = \Gamma_0 \frac{\gamma(\omega)}{\prod\limits_{j=1}^L (\omega-z_j)(\omega-z_j^*)} \,,
\end{align}
where $\Gamma_0>0$ is a constant, $\gamma(\omega)$ is a holomorphic function (or a constant) in the complex plane with $\gamma(\mathbb{R})>0$ and $z_j$ are $L$ different first-order poles of the spectral density in the upper complex half-plane (which implies that $\gamma(z_j) \neq 0$, $\gamma(z_j^*) \neq 0$ and $\Im z_j > 0$).
Since $\Gamma(\omega)\ge 0$ for all $\omega\in\mathbb{R}$, we assume that the first order poles come in complex-conjugate pairs.
In the upper half-plane, $\Gamma(\omega)$ should decay sufficiently fast to enable closure of the contour without the necessity to calculate the corresponding integral contribution.

Then, we get for the principal value integral 
\begin{align}
\bar{\Sigma}(\omega)&=\mathcal{P}\int \frac{d\omega'}{2 \pi} \frac{\Gamma(\omega')}{\omega- \omega'}\nn
&= -\ii \left[\frac{1}{2} \Gamma(\omega)
+ \sum_{j=1}^L \res{\omega'=z_j} \frac{\Gamma(\omega')}{\omega'-\omega}
\right]\nn
&= -\frac{\ii}{2} \Gamma(\omega)\nn
&\qquad - \ii \sum_{j=1}^L \frac{\Gamma_0 \gamma(z_j)}{(z_j-\omega) (z_j - z_j^*) \prod\limits_{i\neq j} (z_j - z_i) (z_j - z_i^*)}\nn
&\equiv -\frac{\ii}{2} \Gamma(\omega) - \sum_{j=1}^L \frac{K_j}{z_j - \omega}\,,
\end{align}
where we have introduced the constant $K_j\in\mathbb{C}$.
Since $\bar{\Sigma}(\omega)\in\mathbb{R}$, we can express the mapping relation~(\ref{EQ:mapping0}) as
\begin{align}
\tilde{\Gamma}(\omega) &= \frac{\lambda^2 \Gamma(\omega)}{\abs{\omega-\epsilon-\bar{\Sigma}(\omega) - \frac{\ii}{2} \Gamma(\omega)}^2}\nn
&= \frac{\lambda^2 \Gamma(\omega)}{\abs{\omega-\epsilon+ \sum\limits_{j=1}^L \frac{K_j}{z_j - \omega}}^2}\,.
\end{align}
Inserting our assumption for the spectral density~(\ref{EQ:spectral_density}), we find that
\begin{align}\label{EQ:appmapped_final}
\tilde{\Gamma}(\omega) &= \frac{\lambda^2 \Gamma_0 \gamma(\omega)}{\abs{
(\omega-\epsilon)\prod\limits_j (\omega-z_j) - \sum\limits_j K_j \prod\limits_{i\neq j} (\omega-z_i)}^2}\nn
&= \frac{\lambda^2 \Gamma_0 \gamma(\omega)}{\prod\limits_{j=1}^{L+1} (\omega-\tilde{z}_j)(\omega-\tilde{z}_j^*)}\,.
\end{align}
Here, the first line reveals that the expression in the absolute value is a polynomial of order $L+1$, which has $L+1$ complex roots, which will have to be determined numerically.
Altogether, one finds $L+1$ pairs of complex conjugate poles $\tilde{z}_j$ for the transformed spectral density.
When these are sufficiently far apart from each other but close to the real axis, every pole pair corresponds to a peak in the spectral density.
This means that by performing the reverse mapping once with such spectral densities, at most a single peak is added
in the transformed spectral density.
Since the resulting shape reproduces the initial assumption~(\ref{EQ:spectral_density}), we can conclude that if one starts with a spectral density
of that form, with each mapping, at most a single peak will be added to the resulting spectral density.

%%%%%%%%%%%%%%%%%%%%%%%%%%%%%%%%%%%%%%%%%%%%%%%%%%%%%%%%%%%%%%%%%%%%%%%%%%%%%%%%%%%%%%%%%%%%%%%%%%%
%%%%%%%%%%%%%%%%%%%%%%%%%%%%%%%%%%%%%%%%%%%%%%%%%%%%%%%%%%%%%%%%%%%%%%%%%%%%%%%%%%%%%%%%%%%%%%%%%%%
%%%%%%%%%%%%%%%%%%%%%%%%%%%%%%%%%%%%%%%%%%%%%%%%%%%%%%%%%%%%%%%%%%%%%%%%%%%%%%%%%%%%%%%%%%%%%%%%%%%
%%%%%%%%%%%%%%%%%%%%%%%%%%%%%%%%%%%%%%%%%%%%%%%%%%%%%%%%%%%%%%%%%%%%%%%%%%%%%%%%%%%%%%%%%%%%%%%%%%%

\section{Computation of the Greens function}\label{APP:greens_function}

To evaluate quantities such as the transmission~\cite{haug2008,economou2006}
\begin{align}
T(\omega) = \abs{G_{1,N}(\omega)}^2 \Gamma_L(\omega) \Gamma_R(\omega)
\end{align}
or the occupation of the systems site $j$
\begin{align}
n_j &= \int \frac{d\omega}{2\pi} \Big[\abs{G_{j,1}(\omega)}^2 \Gamma_L(\omega) f_L(\omega)\nn
&\qquad+ \abs{G_{j,N}(\omega)}^2 \Gamma_R(\omega) f_R(\omega)\Big]\,,
\end{align}
knowledge of the retarded Greens function matrix elements $G_{ij}(\omega)$ is essential.
The full Greens function $G(\omega)$ can be obtained from the free Greens function $G_0(\omega)$ via
solving the Dyson equation
\begin{align}\label{EQ:dyson_solution}
G(\omega) = \left[\f{1} - G_0(\omega) {\cal H}_1\right]^{-1} G_0(\omega)\,.
\end{align}
Here, the free Greens function is based on a splitting of the total (system and reservoir) Hamiltonian into two parts $H = H_0 + H_1$, and ${\cal H}_1$ is the single-particle matrix representation of $H_1$.
When the free Hamiltonian is just a homogeneous chain 
\begin{align}
H_0 &= \sum_{j=-\infty}^{+\infty} \tau_0 \left(c_{j+1}^\dagger c_j + {\rm h.c.}\right)\,,
\end{align}
the retarded free Greens function can be computed analytically and becomes in the single-particle subspace
\begin{align}
G_{0,\ell m}(\omega) = \frac{-\ii}{\sqrt{4\tau_0^2-\omega^2}} \left(\frac{\omega}{2\tau_0} 
-\ii \sqrt{1-\frac{\omega^2}{4\tau_0^2}}\right)^{\abs{\ell-m}}\,.
\end{align}
The central full Greens function's matrix elements can then be computed from~(\ref{EQ:dyson_solution}) with a finite-size matrix inversion if ${\cal H}_1$ is only local.
For example, if the left reservoir sites are labeled from $-\infty < j \le 0$, and the right reservoir sites are labeled from $N+1 \le j < +\infty$, such that the system ranges from sites $1\le j \le N$, and if in addition ${\cal H}_1$ only affects sites $0 \le j \le N+1$, it suffices to invert only the block from $0 \le j \le N+1$, such that an $(N+2)\times(N+2)$ matrix inversion is necessary to compute the central part of $G(\omega)$.
Furthermore, this choice of $H_0$ leads to spectral densities with the shape~(\ref{EQ:specdens_semicircle}).
When $\tau_\alpha = \sqrt{\Gamma_\alpha \tau_0/2}$, this resembles in the limit $\tau_0\to\infty$ a flat spectral density as used for comparisons in the main text.

%%%%%%%%%%%%%%%%%%%%%%%%%%%%%%%%%%%%%%%%%%%%%%%%%%%%%%%%%%%%%%%%%%%%%%%%%%%%%%%%%%%%%%%%%%%%%%%%%%%
%%%%%%%%%%%%%%%%%%%%%%%%%%%%%%%%%%%%%%%%%%%%%%%%%%%%%%%%%%%%%%%%%%%%%%%%%%%%%%%%%%%%%%%%%%%%%%%%%%%
%%%%%%%%%%%%%%%%%%%%%%%%%%%%%%%%%%%%%%%%%%%%%%%%%%%%%%%%%%%%%%%%%%%%%%%%%%%%%%%%%%%%%%%%%%%%%%%%%%%
%%%%%%%%%%%%%%%%%%%%%%%%%%%%%%%%%%%%%%%%%%%%%%%%%%%%%%%%%%%%%%%%%%%%%%%%%%%%%%%%%%%%%%%%%%%%%%%%%%%
\section{Numerical calculation of the mapping}

\subsection{Straight-line interpolation}\label{APP:pval1}

For a representation of the tunneling rate $\Gamma(\omega)$ in the interval $[\omega_{\rm min}, \omega_{\rm max}]$
by straight-line approximations between ordered ($\omega_i < \omega_{i+1}$) sampling points $\omega_{\rm min} \le \omega_i \le \omega_{\rm max}$ we
can -- abbreviating $\Gamma_i = \Gamma(\omega_i)$ -- write for $\omega \in [\omega_i, \omega_{i+1}]$
\begin{align}
\Gamma(\omega) = \Gamma_i + (\Gamma_{i+1}-\Gamma_i) \frac{\omega-\omega_i}{\omega_{i+1}-\omega_i}\,.
\end{align}
For such a straight-line interpolation, the principal value integral
\begin{align}\label{EQ:integral_type}
I_{i,i+1}(\omega) = {\cal P} \int_{\omega_i}^{\omega_{i+1}} \frac{\Gamma(\omega')}{\omega-\omega'} d\omega'
\end{align}
can be computed analytically, provided that appropriate case distinctions are performed.
When $\omega$ is outside the interval ($\omega < \omega_i$ or $\omega>\omega_{i+1}$), the integral becomes an ordinary one 
\begin{align}\label{EQ:int1}
I_{i,i+1}(\omega) \to 
(\Gamma_i - \Gamma_{i+1}) +&  \frac{\Gamma_i (\omega_{i+1}-\omega) + \Gamma_{i+1}(\omega-\omega_i)}{\omega_{i+1}-\omega_i} \nn
 & \, \, \times \ln \left[\frac{\omega_i-\omega}{\omega_{i+1}-\omega}\right]\,.
\end{align}
When $\omega$ is inside the interval ($\omega_i < \omega < \omega_{i+1}$), this expression only changes slightly
\begin{align}\label{EQ:int2}
I_{i,i+1}(\omega) \to (\Gamma_i - \Gamma_{i+1})
+ & \frac{\Gamma_i (\omega_{i+1}-\omega) + \Gamma_{i+1}(\omega-\omega_i)}{\omega_{i+1}-\omega_i} \nn
 & \, \, \times \ln \left[\frac{\omega-\omega_i}{\omega_{i+1}-\omega}\right]\,.
\end{align}
Finally, when $\omega=\omega_i$, we have to combine the contributions from intervals $[\omega_{i-1},\omega_i]$ and $[\omega_i,\omega_{i+1}]$
to appropriately balance the divergent contributions
\begin{align}\label{EQ:int3}
I_{i-1,i}(\omega_i)+ & I_{i,i+1}(\omega_i) \to \nn
& (\Gamma_{i-1}-\Gamma_{i+1})
+ \Gamma_i \ln \left[\frac{\omega_i-\omega_{i-1}}{\omega_{i+1}-\omega_i}\right]\,,
\end{align}
which however cannot be applied at the boundaries of the discretization (i.e., $i>1$ and $i<N_d$).
There, under the assumption that $\Gamma_1 = \Gamma(\omega_{\rm min}) = 0 = \Gamma(\omega_{\rm max}) = \Gamma_{N_d}$, we obtain the
simplified contributions
$I_{1,2} \to -\Gamma_2$ and $I_{N-1,N} \to +\Gamma_{N-1}$.
For a numerical implementation of integrals of the type~(\ref{EQ:integral_type}), we have to sum up the contributions from the individual discretization
intervals, where it becomes obvious that the terms in round brackets in Eqns.~(\ref{EQ:int1}), (\ref{EQ:int2}), and~(\ref{EQ:int3}) will cancel with the boundary contributions such that only the $\ln[\ldots]$ terms need to be considered.
%%%%%%%%%%%%%%%%%%%%%%%%%%%%%%%%%%%%%%%%%%%%%%%%%%%%%%%%%%%%%%%%%%%%%%%%%%%%%%%%%%%%%%%%%%%%%%%%%%%

\subsection{Lorentzian approximation}\label{APP:pval2}

With the previous approach, a problem may arise when the peaks in the mapped spectral densities become very narrow, in particular for higher order mappings.
This causes the need for more gridpoints to resolve the peaks, which then increases the computational effort. 
To avoid this issue, one can fit peaked spectral densities by sums of Lorentzian functions.
Given $m$ peaks in the spectrum, we fit each with the function
\begin{align}\label{EQ:fit}
\tilde{\Gamma}_k(\omega)=\frac{\tilde{\Gamma}_k \delta_k^2}{(\omega-\epsilon_k)^2+\delta_k^2}\, , \,\,\,\,\,\,\,\,\,\,\,\,\, k=1,\ldots,m \,.
\end{align}
This has its origin in the fact that flat spectral densities can be mapped to functions of this form (see the main text). 
Technically, such an approximation has the advantage that 
the principal value integrals can be easily performed
\begin{align}
\bar{\Sigma}_k(\omega)=\int \frac{d\omega'}{2\pi} \frac{\tilde{\Gamma}_k(\omega')}{\omega-\omega'}=\frac{\tilde{\Gamma}_k \delta_k (\omega - \epsilon_k)}{2 (\delta_k^2 + (\omega - \epsilon_k)^2)}\,.
\end{align}
This allows one to easily calculate the mapping relation~(\ref{EQ:mapping}).

\begin{figure}[ht]
\includegraphics[width=0.45\textwidth,clip=true]{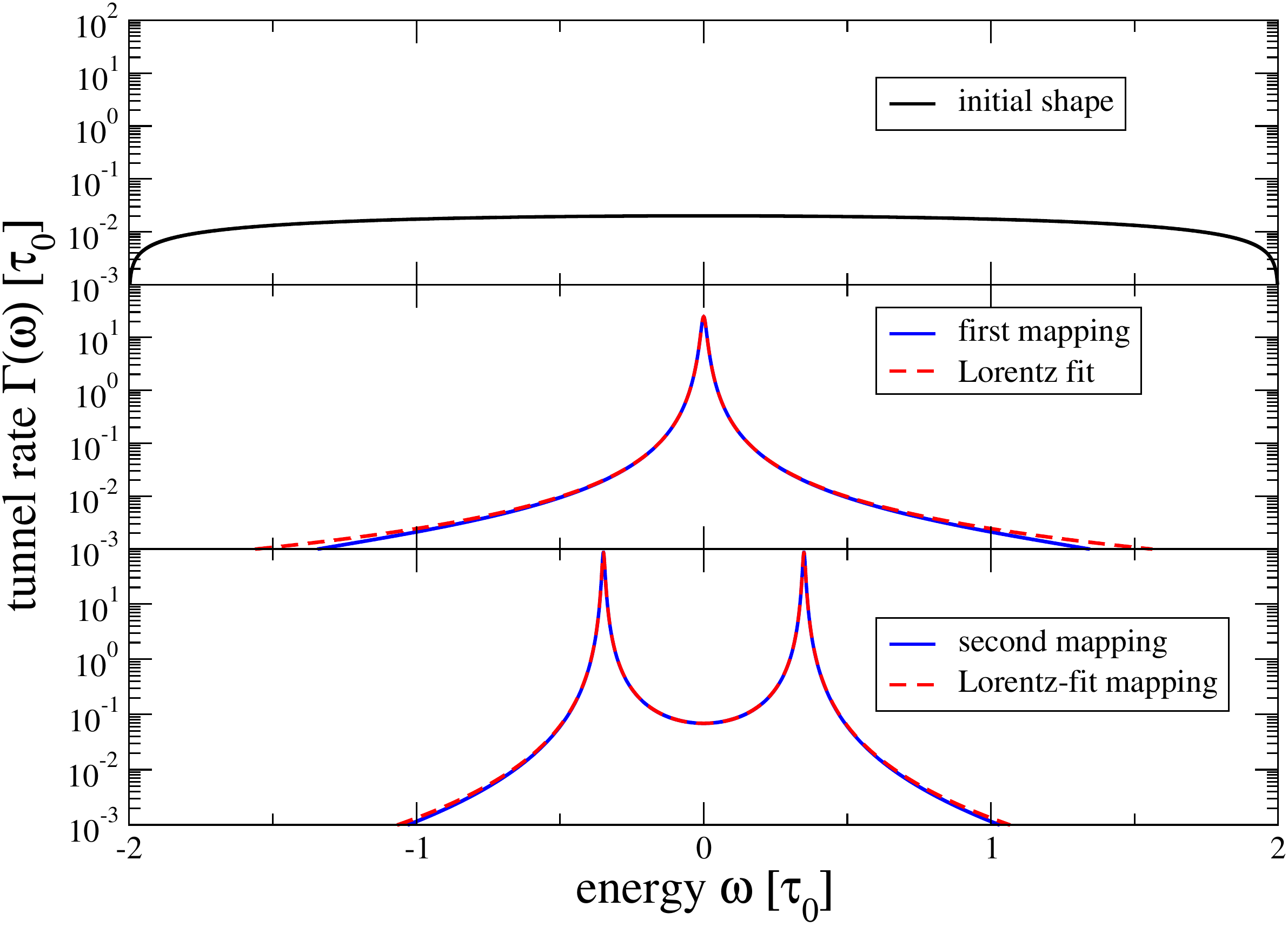}
\caption{\label{FIG:mappingExample}
First few spectral densities for the SSH-chain transmission in Fig.~\ref{FIG:transmissionssh} with parameters as given there.
In the top panel, the initial spectral density is shown (note the logarithmic scale).
In the middle panel we provide the first mapping as an exact calculation (solid blue) and a Lorentz fit of the data (dashed red), which
agree reasonably well over four orders of magnitude.
In the bottom panel the second mapping calculated by numerical integration is shown (solid blue) and with the Lorentzian fit (dashed red).
}
\end{figure}

Even if the original spectral density is not flat, such Lorentzian shapes are still found approximately, 
requiring that the peaks in the spectral densities are well separated. 
We exemplify this in Fig.~\ref{FIG:mappingExample}.
In the top panel, the initial spectral density (identical for both reservoirs) is shown, 
which we used for the calculation for the SSH model in Fig.~\ref{FIG:transmissionssh}. 
The first mapping is shown in the middle panel. 
The solid line is the exact analytic solution for a semicircle spectral density, while the dashed is a fit of the peak with the Lorentz from
Eq.~(\ref{EQ:fit}). 
One can see that the fit approximates the analytic solution in the important central region well. 
This also holds for higher order mappings.
In Fig.~\ref{FIG:mappingExample} bottom we show the second mapping calculated via numerical integration as the solid line, while the dashed curve shows the analytic mapping of the fitted Lorentzian. 
Also these results fit well over orders of magnitude.
If the tunneling rate is not symmetric around the peak, by using a Lorentz fit the information on the asymmetry is lost.
This causes that both peaks in the next mapping will have the same height, while the calculation via direct numerical integration shows that the peak heights actually vary.
To use the advantage of the fit without loosing information, we suggest to combine Lorentz fitting 
with brute-force numerical integration of the residual spectral density:
For a given peaked spectral density, we fit the peaks with Lorentzian functions as in equation \eqref{EQ:fit}.
If the initial spectral density has a hard cutoff, such that outside a certain interval $[\omega_{\rm min},\omega_{\rm max}]$ the spectral density strictly vanishes, 
an analytic representation of the principal value integrals can be found as well.
For example, one can approximate each peak by fitting a Lorentzian function and continue it after some
cutoff until it reaches zero, e.g. 
\begin{align}
\hat{\Gamma}_k(\omega)=\left\{
	\begin{array}{ccc}
		+\Delta (\omega - \omega_L) & : &\omega_L \le \omega < \epsilon_k-c\\
    		\tilde{\Gamma}_k(\omega) & : & \epsilon_k-c \le \omega < \epsilon_k+c \\
    		-\Delta (\omega - \omega_R) &  : & \epsilon_k+c \le \omega < \omega_R \\
   		0 & : & \textrm{else}
	\end{array}
\right. \,,
\end{align}
where $c>0$ is a cutoff, such that $\hat{\Gamma}_k(\omega) = \tilde{\Gamma}_k(\omega)$ for 
$\omega \in [\epsilon_k-c,\epsilon_k+c]$. 
The slope $\Delta$ and the frequencies $\omega_{L/R}$ are not independent.
They have to be chosen such that at $\omega = \epsilon_k\pm c$, both the function and its derivative are continuous, which yields
$\Delta = \frac{2 \Gamma_k \delta_k^2 c}{(\delta_k^2 + c^2)^2}$ and 
$\omega_{L/R}=\epsilon_k \mp\frac{3}{2} c \mp \frac{\delta_k^2}{2 c}$.
This parametrization has the advantage that the principal value integral in the mapping relation will lead to a continuous function, and by choosing the cutoff $c$ carefully, the function takes nonzero values only in the interval 
$[\omega_L,\omega_R]\subseteq [\omega_{\rm min},\omega_{\rm max}]$.
The fit of all peaks is then defined as $\hat{\Gamma}(\omega)=\sum_{k=1}^m \hat{\Gamma}_k(\omega)$ and the level renormalization is given by $\hat{\Sigma}(\omega) = \sum_{k=1}^m \hat{\Sigma}_k (\omega)$, which is analytically solvable.

After fitting the peaks we create a sampling in the Interval $\left[\omega_{\rm min}, \omega_{\rm max} \right]$ and calculate the residual spectral density, i.e., the difference between the original data and the fit. 
Importantly, the residual spectral density strictly vanishes at $\omega_{\rm min}$ and $\omega_{\rm max}$, such that the direct integration
from the previous section can be employed.
By the linearity of the integral, the level renormalization will then just be a sum of the renormalizations of the Lorentz fit and the residual spectral density.
The advantage of this procedure is that the residual spectral density is smoother/flatter, such that a coarser discretization grid can be used.

\subsection{Semi-analytic calculation}\label{APP:pval3}

\begin{figure}[ht]
\includegraphics[width=0.45\textwidth,clip=true]{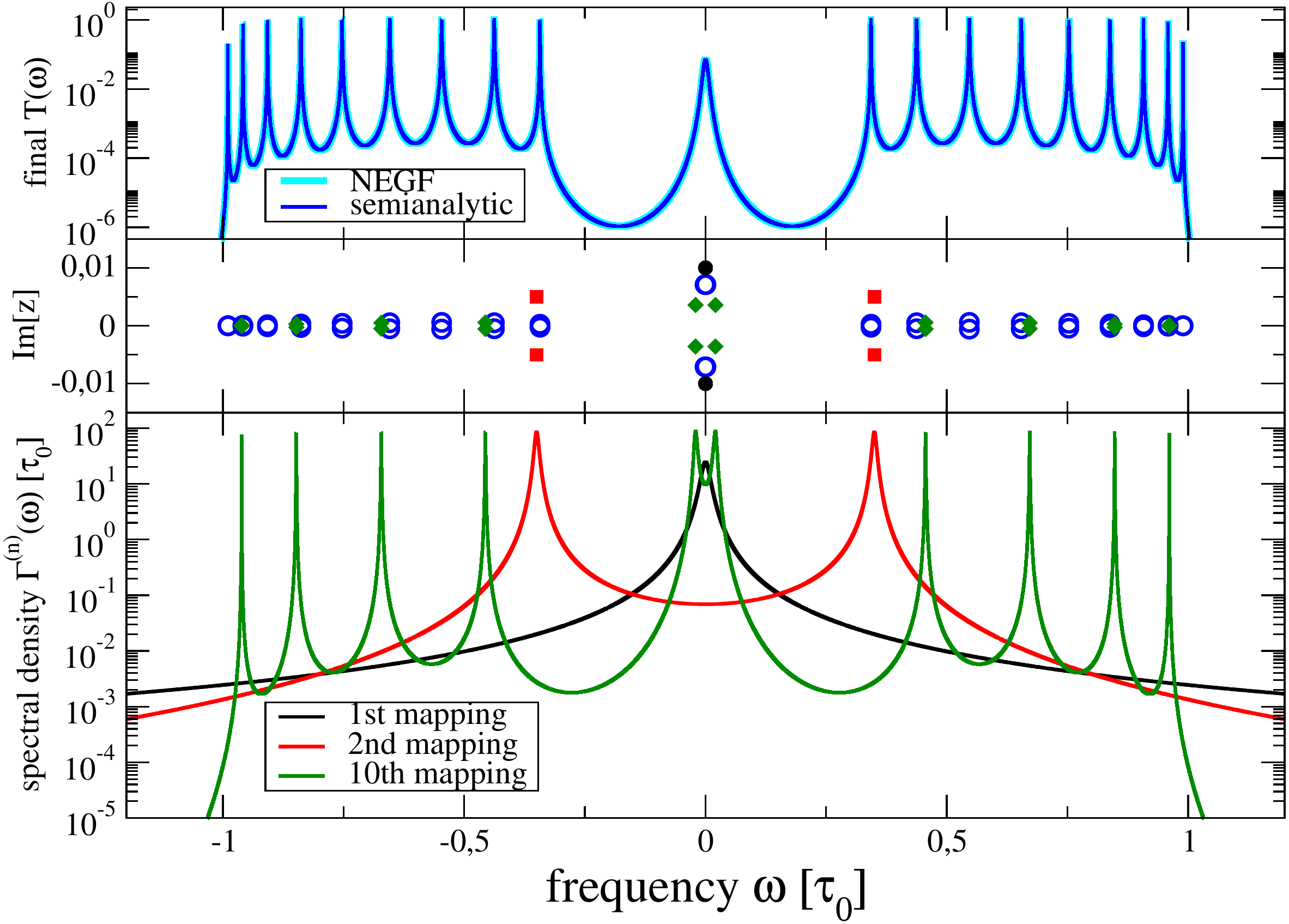}
\caption{\label{FIG:polestruct}
Plot of the semianalytically obtained spectral densities for 10 mappings of the SSH model (bottom).
Peaks of the spectral density are aligned with complex-conjugate pole pairs in the complex plane (middle).
For an SSH chain of $20$ sites, one can use 9 mappings to the left and 10 mappings to the right reservoir. 
Combining these yields a transmission (top) described by 20 pole pairs in the complex plane (middle, blue), which agrees perfectly with the non-equilibrium Greens function calculation when the semicircle spectral density is nearly flat (light blue). 
Parameters have been chosen as in Fig.~\ref{FIG:transmissionssh}, with the exception that the initial spectral density was assumed flat, matching the maximum of the semicircle spectral density~(\ref{EQ:specdens_semicircle}) with $\Gamma_0=\Gamma^{(20)}(0) = 0.02\tau_0$.
}
\end{figure}

Here, we outline how to compute the reverse reaction-coordinate mapping for spectral densities that are initially flat.
From Eq.~(\ref{EQ:appmapped_final}) we are thus facing the problem to compute the complex roots of the polynomial
\begin{align}
P(\omega) = (\omega-\epsilon)\prod\limits_{j=1}^L (\omega-z_j) - \sum\limits_{j=1}^L K_j \prod\limits_{i\neq j} (\omega-z_i)\,.
\end{align}
The total pole structure is given by these roots and the complex conjugate ones.
Here, $\epsilon$ is the on-site energy of the boundary dot, $L$ is the number of poles $z_i : \Im(z_i) > 0$ in the upper complex half plane of the previous spectral density, and
\begin{align}\label{EQ:kvec}
K_j = \ii \frac{\Gamma_0 \gamma(z_j)}{(z_j - z_j^*) \prod\limits_{i\neq j} (z_j - z_i) (z_j - z_i^*)}\,.
\end{align}
Particularly, when one starts from an initially flat spectral density $\Gamma_0$, one has to set 
$\gamma(z_j)=1=\gamma(\omega)$.
We can take the known old poles $z_i$ together with the onsite-energy $\epsilon$ -- supplemented with a small imaginary 
component -- as initial guesses for the 
iterative Weierstra{\ss}-Durand-Kerner algorithm
\begin{align}
\tilde{z}_i^{(n+1)} = \tilde{z}_i^{(n)} - \frac{P(\tilde{z}_i^{(n)})}{\prod\limits_{j\neq i} (\tilde{z}_i^{(n)} - \tilde{z}_j^{(n)})}\,,
\end{align}
which has to be iterated until convergence.
Practically, we found for our parameters rapid convergence to machine accuracy after only roughly $N$ iterations for a polynomial of order $N$.
In Fig.~\ref{FIG:polestruct} one can see that the found poles in the complex plane (middle panel) match the peaks in the spectral density (bottom panel).
Therefore, given an electronic transport problem with flat spectral densities, we would consider the semianalytic mapping as a very useful approach.

We note also that the calculation of the transmission of the residual single dot from the structured residual spectral densities can be performed 
along the same lines:
Having determined the constants $K_{j,L/R}$ from Eq.~(\ref{EQ:kvec}) individually for the residual spectral densities $\Gamma_L(\omega)$ and $\Gamma_R(\omega)$ with $N_\alpha$ poles $z_{j,\alpha}$ in the upper complex half plane, the transmission of the remainder dot~(\ref{EQ:transmission}) can be written as
\begin{align}
T(\omega) = \frac{\Gamma_0^L \Gamma_0^R}{\prod\limits_i(\omega-z_i)(\omega-z_i^*)}\,,
\end{align}
where the poles of the transmission in the complex plane $z_i$ are given by the roots of the polynomial
\begin{align}
P(\omega) &= (\omega-\epsilon) \left[\prod_{j=1}^{N_L} (\omega - z_{j,L})\right]\left[\prod_{k=1}^{N_R} (\omega - z_{k,R})\right]\nn
&\qquad- \sum_{i=1}^{N_L} K_{i,L} \left[\prod_{j\neq i} (\omega-z_{j,L})\right] \left[\prod_{k=1}^{N_R} (\omega-z_{k,R})\right]\nn
&\qquad- \sum_{i=1}^{N_R} K_{i,R} \left[\prod_{j=1}^{N_L} (\omega-z_{j,L})\right] \left[\prod_{k\neq i} (\omega-z_{k,R})\right]\,.
\end{align}
Here, $\epsilon$ is the on-site energy of the remainder dot.
The resulting transmission will then have $N_L+N_R+1$ pole pairs in total.
Particularly, combining e.g. the $9$th mapping to the left reservoir and the $10$th mapping to the right reservoir, we obtain the transmission of an SSH chain of $20$ sites, described by $20$ pole pairs in the complex plane, compare top panel and hollow blue symbols in the middle panel of Fig.~\ref{FIG:polestruct}.
This transmission agrees perfectly with the independently calculated transmission for an SSH chain with 20 sites with nonequilibrium Greens functions (light blue), where the initially flat bands were simulated from Eq.~(\ref{EQ:specdens_semicircle}) by $\tau_\alpha = \sqrt{\Gamma_\alpha \tau_0/2}$ 
and then considering $\tau_0\to\infty$.

\end{document}